\documentstyle[psfig]{l-aa}
\let\ov=\over
\let\l=\left
\let\r=\right
\def \der#1#2{{\partial{#1}\over\partial{#2}}}
\def \dder#1#2{{\partial^2{#1}\over\partial{#2}^2}}
\def\be{\begin{equation}}
\def\ee{\end{equation}}

\begin{document}

\thesaurus{06(02.18.8; 08.13.1; 08.14.1; 08.16.6; 08.18.1; 03.13.4)}

\title{Rotating neutron star models with magnetic field}

\author{M.~Bocquet\inst{1,2} \and S.~Bonazzola\inst{1}
\and E.~Gourgoulhon\inst{1}\fnmsep\thanks{author to whom the proofs should be
sent} \and J.~Novak\inst{1,2} }

\offprints{E.~Gourgoulhon}

\institute{D\'epartement d'Astrophysique Relativiste et de Cosmologie
  (UPR 176 C.N.R.S.), Observatoire de Paris, \\
   Section de Meudon, F-92195 Meudon Cedex, France \\
   {\em e-mail : gourgoulhon@obspm.fr}
 \and
    Ecole Polytechnique, F-91128 Palaiseau Cedex, France  }

\date{Received date; accepted date}

\maketitle

\begin{abstract}
We present the first numerical solutions of the coupled Einstein-Maxwell
equations describing rapidly rotating neutron stars endowed with a
magnetic field. These solutions are fully relativistic and
self-consistent, all the effects of the
electromagnetic field on the star's equilibrium (Lorentz force, spacetime
curvature generated by the electromagnetic stress-energy) being taken into
account. The magnetic field is axisymmetric and poloidal. Five dense matter
equations of state are employed. The partial differential equation system
is integrated by means of a pseudo-spectral method.
Various tests passed by the numerical code are presented. The effects of
the magnetic field on neutron stars structure are then investigated,
especially by comparing magnetized and non-magnetized configurations with the
same baryon number. The deformation of the star induced by the magnetic field
is important only for huge values of $\vec{B}$ ($B>10^{10} {\ \rm T}$).
The maximum mass as well as the maximum rotational velocity are found to
increase with the magnetic field. The maximum allowable poloidal magnetic field
is of the order of $10^{14} {\ \rm T}$ and
is reached when the magnetic pressure is comparable to the fluid pressure at
the
centre of the star. For such values, the maximum mass of neutron stars
is found to increase by $13$ to $29\%$ (depending upon the EOS) with
respect to the maximum mass of non-magnetized stars.
\end{abstract}

\section{Introduction} \label{s:intro}

Neutron stars are known to possess strong magnetic fields. Their polar
field strength is deduced from the observed spin slowdown of pulsars via
the magnetic dipole braking model;
for the 558 pulsars of the catalog by Taylor et al. (1993), it ranges
from $B=1.7\ 10^{-5}$ GT \footnote{In this article,
we systematically use S.I. units, so that the magnetic field amplitude
is measured in teslas (T) or, more conveniently in gigateslas ($1{\ \rm GT} =
10^9{\ \rm T}$). We recall that $1{\ \rm GT} =  10^{13}$ gauss.}
(PSR B1957+20) up to $B=2.1$ GT (PSR B0154+61), with a median value $B=0.13$
GT,
most of young pulsars having a surface field in the range
$B\sim 0.1 - 2 {\ \rm GT}$.
{}From the theoretical point of
view, a considerable amount of studies have been devoted to the structure of
the magnetic field {\em outside} the neutron star, in the so-called
{\em magnetosphere}, in relation with the pulsar emission mechanism
(for a review, see e.g. Michel 1991). The studies of the
magnetic field {\em inside} neutron stars are far less abundant.
Only recently, some works have been devoted to the origin and the
evolution of the internal magnetic field, all of them in the non-relativistic
approximation (Thompson \& Duncan 1993, Urpin \& Ray 1994,
Wiebicke \& Geppert 1995, Urpin \& Shalybkov 1995).

Beside these studies of neutron star magnetic field, there exists a
growing number of numerical computations of rapidly rotating neutron stars
in the full framework of general relativity, taking into account
the most sophisticated equations of state of dense matter to date
(cf. Salgado et al.~1994a,b and reference therein, as well as
Cook et al.~1994b, Eriguchi et al.~1994,
Friedman \& Ipser 1992). But in all these models
the magnetic field is ignored. The present work is the first attempt
to compute numerical models of rotating neutron stars with magnetic
field in a self-consistent way, by solving the Einstein-Maxwell
equations describing stationary axisymmetric rotating objects with
internal electric currents. In this way, the models presented below
\begin{enumerate}
\item are fully relativistic, i.e. all the effects of general relativity
are taken into account, on the gravitational field as well as on the
electromagnetic field.
\item are self-consistent, i.e. the electromagnetic field is generated
by some electric current distribution and
the equilibrium of the matter is
given by the balance between the gravitational force, the pressure
gradient and the Lorentz force corresponding to the electric current.
Moreover, the electromagnetic energy density is taken into account in the
source of the gravitational field.
\item give the solution in all space, from the star's centre to infinity,
without any approximation on the boundary conditions.
\item use various equations of state proposed in the literature for
describing neutron star matter.
\end{enumerate}
The restrictions of our models are the following ones:
\begin{enumerate}
\item We consider strictly stationary configurations. This excludes magnetic
dipole moment non-aligned with the rotation axis. Indeed, in the
non-aligned case, the star radiates away electromagnetic waves as well
as gravitational waves (due to the deviation from axisymmetry induced
by a magnetic axis different from the rotation axis); hence it loses
energy and angular momentum, so that this situation does not correspond
to any stationary solution of Einstein equation.  Thus the stationary
hypothesis implies that we restrict ourselves to axisymmetric
configurations, with the magnetic axis aligned with the rotation axis.

\item Moreover, we consider only {\em poloidal} magnetic fields (i.e. $\vec{B}$
lying in the meridional planes). Indeed, if the magnetic field had, in
addition to the poloidal part, a {\em toroidal} component
(i.e. a component perpendicular to the meridional planes), the
{\em circularity} property of spacetime would be broken, which means that the
two Killing vectors associated with
the stationarity and the axisymmetry would no longer be orthogonal to a family
of 2-surfaces (cf. Carter 1973, p.~159). In the circular case, a coordinate
system $(t,r,\theta,\phi)$ can be found such that the components of the metric
tensor $\vec{g}$ are zero except for the diagonal terms and only one
off-diagonal term ($g_{t\phi}$). In the non-circular case, only one component
of $\vec{g}$ can be set to zero ($g_{r\theta}$), resulting in much more
complicated gravitational field equations (Gourgoulhon \& Bonazzola 1993).
On the contrary,
perfect fluid stars with purely rotational motion (no convection) generate
circular spacetimes and poloidal magnetic fields preserve this property
(Carter 1973).

\item Being not interested in modelling pulsar magnetospheres, we
suppose that the neutron star is surrounded by vacuum\footnote{by {\em vacuum}
we mean that there is no matter outside the star; nevertheless there
is some electromagnetic field, so that the total stress-energy tensor
(right-hand side of the Einstein equation) is not zero outside the star
({\em electrovac} spacetime).}.
\end{enumerate}

The numerical code we use is an electromagnetic extension of the code
presented in Bonazzola et al.~1993 (hereafter BGSM), which was devoted
to perfect fluid rotating stars and used to compute neutron star models
with various equations of state of dense matter (Salgado et
al.~1994a,b).  We will not give here the complete list of the equations
to be solved but only the electromagnetic ones (Maxwell equations) in
Sect.~\ref{s:electromagn}, referring to BGSM for the gravitational
part.  Likewise we will not present the numerical technique, based on a
pseudo-spectral method, since it has been detailed in BGSM. We will
discuss in Sect.~\ref{s:tests} only the numerical procedure and tests
of the electromagnetic part of the code. We analyze the effects of the
magnetic field on static configurations in Sect.~\ref{s:static} and
on rotating configurations in Sect.~\ref{s:rotat},
the Sect.~\ref{s:const,bar} describing in detail the case of constant baryon
number sequences. Finally
Sect.~\ref{s:concl} summarizes the main conclusions of this study.

\section{Electromagnetic equations} \label{s:electromagn}

\subsection{Definitions and notations} \label{s:def,not}

Following BGSM, we use
MSQI ({\em Maximal Slicing - Quasi-Isotropic}) coordinates $(t,r,\theta,\phi)$,
in which the metric tensor $\vec{g}$ takes the form
\begin{eqnarray}
g_{\alpha\beta} \, dx^\alpha \, dx^\beta  = &
 - N^2 \, dt^2 + A^4 \Big[ B^2 r^2 \sin^2 \theta
  \left( d\phi-N^\phi dt \right) ^2  \nonumber \\
  & \displaystyle
  + \frac{1}{B^2} \left( dr^2 + r^2 d\theta^2 \right) \Big] \ ,
\end{eqnarray}
where $N$, $N^\phi$, $A$ and $B$ are four functions of $(r,\theta)$, sometimes
represented by their logarithms:
\be \label{e:def:nu,a,b}
  \nu := \ln N \quad ; \quad \alpha := \ln A \quad ; \quad
	\beta := \ln B \ .
\ee

Carter (1973, theorem 7) has shown that the most general form of the electric
4-current $\vec{j}$ compatible with the hypothesis of stationarity,
axisymmetry and circularity has the following components with respect
to the $(t,r,\theta,\phi)$ coordinates:
$j^\alpha = (j^t,0,0,j^\phi)$.
He has also shown that the corresponding electromagnetic field tensor
$\vec{F}$ must be
derived from a potential 1-form $\vec{A}$ with the following components
$A_\alpha = (A_t,0,0,A_\phi)$:
\be \label{e:F=dA}
   F_{\alpha\beta} = A_{\beta,\alpha} - A_{\alpha,\beta} \ .
\ee
We will not use the ``orthonormal''
electromagnetic potentials $\Phi$ and $\Psi$ introduced in BGSM but work
directly with the components $A_t$ and $A_\phi$.
A nice feature of $A_\phi$ is that the magnetic field lines
lie on the surfaces $A_\phi={\rm const}$.

For our problem, a priviledged observer is the observer ${\cal O}_0$
whose 4-velocity $\vec{n}$
is the unit vector normal to the $t={\rm const}$ hypersurfaces
(hereafter referred to as the $\Sigma_t$ hypersurfaces): it corresponds to
the observer ``at rest with respect to the star's centre''
of the Newtonian physics.
It is called the {\em locally nonrotating observer} by Bardeen (1970),
the {\em zero-angular-momentum observer (ZAMO)} by Bardeen (1973), the
{\em Eulerian observer} by Smarr \& York (1978) and the {\em FIDO} by
Thorne et al. (1986). The electric field $\vec{E}$ and the
magnetic field $\vec{B}$
as measured by the observer ${\cal O}_0$ are given by\footnote{Unless otherwise
specified, we will refer in the
following to $\vec{E}$ and $\vec{B}$ as the electric field and the
magnetic field, without mentioning ``as measured by the observer
${\cal O}_0$''.}
\begin{eqnarray} \label{e:E_alpha}
    E_\alpha & = & F_{\alpha\beta} n^\beta  \nonumber \\
     & = & \l(\! 0 , {1\ov N} \l[ \der{A_t}{r} + N^\phi \der{A_\phi}{r} \r] ,
    {1\ov N} \l[ \der{A_t}{\theta} + N^\phi \der{A_\phi}{\theta} \r] , 0 \! \r)
		\ ;
\end{eqnarray}
\begin{eqnarray} \label{e:B_alpha}
   B_\alpha & = &
     - {1\ov 2} \, \epsilon_{\alpha\beta\mu\nu} \, n^\beta \,  F^{\mu\nu}
				\nonumber \\
    & = & \l( 0 , {1\ov A^2 B r^2\sin\theta} \der{A_\phi}{\theta},
           - {1\ov A^2 B \sin\theta} \der{A_\phi}{r}, 0 \r)    \ ,
\end{eqnarray}
where $\epsilon_{\alpha\beta\mu\nu}$ is the Levi-Civita tensor associated with
the metric $\vec{g}$.
Note that a consequence of Eq.~(\ref{e:B_alpha}) is
that the magnetic field lines lie on surfaces $A_\phi = {\rm const}$.
In non-relativistic studies, $A_\phi$ is usually called the
{\em magnetic stream function} or {\em magnetic flux function}
and is denoted $a$ or $A$ (see e.g. Sakurai 1985, Sauty \& Tsinganos 1994).

\subsection{Maxwell equations}

The source-free Maxwell equations $F_{[\alpha\beta;\gamma]}=0$ are
automatically satisfied by the form (\ref{e:F=dA}) of $\vec{F}$.
The remaining Maxwell equations
$F^{\alpha\beta}_{\ \ \  ;\beta} = \mu_0 \, j^\alpha$ can be expressed
in terms of $A_t$ and $A_\phi$ as the Maxwell-Gauss equation
\begin{eqnarray}
   \Delta_3 A_t & = & - \mu_0 {A^4\ov B^2} \l( g_{tt} \, j^t
	+ g_{t\phi}\, j^\phi \r) - {A^4 B^2\ov N^2} N^\phi r^2\sin^2\theta
	  \times \nonumber \\
    & & \times \partial A_t \, \partial N^\phi
	- \l( 1 + {A^4 B^2\ov N^2} r^2\sin^2\theta (N^\phi)^2 \r)
	  \times \nonumber \\
    & & \times \partial A_\phi \, \partial N^\phi -
	( \partial A_t + 2 N^\phi \partial A_\phi ) \partial (2\alpha +
	\beta - \nu) \nonumber \\
     & & - 2 {N^\phi \ov r} \l( \der{A_\phi}{r} + {1\ov r \tan\theta}
		\der{A_\phi}{\theta} \r) \label{e:MaxGau}
\end{eqnarray}
and the Maxwell-Amp\`ere equation
\begin{eqnarray}
   \tilde \Delta_3 {\tilde A}^\phi & = & - \mu_0 \, A^8 \, (j^\phi - N^\phi
j^t)
	\, r \sin\theta \nonumber \\
   & &  + {A^4 B^2\ov N^2} \, r \sin\theta \, \partial N^\phi \,
 	(\partial A_t + N^\phi \partial A_\phi ) \nonumber \\
   & & + {1\ov r\sin\theta} \, \partial A_\phi \, \partial
	( 2\alpha + \beta - \nu) \ , \label{e:MaxAmp}
\end{eqnarray}
where
\be
   \tilde A^\phi := {A_\phi \ov r \sin\theta} \ ,
\ee
use has been made of the abridged notation
\be
    \partial \alpha \, \partial \beta := \der{\alpha}{r} \der{\beta}{r}
    + {1\ov r^2} \der{\alpha}{\theta} \der{\beta}{\theta}  \ ,
\ee
and
$\Delta_3$ and $\tilde \Delta_3$ are, respectively,
the scalar Laplacian and the $\phi$
component of the vector Laplacian, in three dimensional flat
space:
\begin{eqnarray}
   \Delta_3 & := &
    \dder{}{r} + {2\ov r} \der{}{r}
    +{1\ov r^2} \dder{}{\theta} + {1\ov r^2 \tan\theta} \der{}{\theta} \\
   \tilde \Delta_3 & := &
    \dder{}{r} + {2\ov r} \der{}{r}
    +{1\ov r^2} \dder{}{\theta} + {1\ov r^2 \tan\theta} \der{}{\theta}
    - {1 \ov r^2\sin^2\theta} \ .
\end{eqnarray}

\subsection{Lorentz force and condition for a stationary motion}

A first integral of the equation of fluid stationary motion (momentum
conservation equation), taking into account the Lorentz force
exerted by the electromagnetic field on the conducting medium
has been derived in BGSM (Eq.~(5.30)); it reads
\be
   H(r,\theta) + \nu(r,\theta) - \ln \Gamma(r,\theta) + M(r,\theta) =
	{\rm const.} \ ,
\ee
where
\begin{itemize}
\item $H$ is the fluid log-enthalpy defined as
\be \label{e:def:H}
    H := \ln \left( {e+p \over n \, m_{\rm B} c^2 } \right) \ ,
\ee
$e$ being the fluid proper energy density, $p$ the pressure, $n$ the
proper baryon density, $m_{\rm B}$ a mean baryon rest mass
($m_{\rm B} = 1.66\ 10^{-27}$ kg);
\item $\nu$ is the gravitational potential defined by Eq.~(\ref{e:def:nu,a,b});
\item $\Gamma$ is the Lorentz factor relating the observer ${\cal O}_0$ and
the fluid comoving observer (cf. Eq.~(5.6) of BGSM);
\item $M$ is the electromagnetic term induced by the Lorentz force.
\end{itemize}
$M$ is actually a function of $A_\phi$ only, expressible as
\be
      M(r,\theta) = M(A_\phi(r,\theta))
		 = - \int_0^{A_\phi(r,\theta)} f(x) \, dx     \ ,
\ee
where $f$ is an arbitrary regular function relating the components of
the electric current to the electromagnetic potential $A_\phi$ as an
integrability condition of the equation of fluid stationary motion
(cf. Eq.~(5.29) of BGSM):
\be \label{e:jphi=f(Aphi)}
   j^\phi - \Omega j^t  = (e+p) f(A_\phi)   \ ,
\ee
where $\Omega$ is the fluid angular velocity defined as
$\Omega := u^\phi / u^t$. We will call $f$ the {\em current function}.
Different choices for $f$ will lead to different magnetic field distributions.
Note that in the limit of an incompressible Newtonian body, the integrability
condition (\ref{e:jphi=f(Aphi)}) reduces to the relation derived by
Ferraro (1954) [cf. his Eqs.~(11) and (14), his function $U$ being exactly
our $A_\phi$ and his function $f$ being $-\mu_0 \rho c^2$ times our
function $f$].

\subsection{Perfect conductor relation}

According to Ohm's Law, and assuming that the matter has an infinite
conductivity, the electric field as measured by the fluid comoving observer,
$E'_\alpha = F_{\alpha\beta} u^\beta$, must be zero.  This condition
leads to the following relation between the two components of the
potential 4-vector inside the star:
\be \label{e:dAt=-Omega dAphi}
   \der{A_t}{x^i} = - \Omega \der{A_\phi}{x^i}             \ .
\ee
As shown in BGSM, a stationary configuration with a magnetic field is
necessarily {\em rigidly} rotating (i.e. has $\Omega={\rm const.}$).
Consequently, Eq.~(\ref{e:dAt=-Omega dAphi}) is integrated immediately
to
\be \label{e:conductparfait}
   A_t = - \Omega A_\phi + C \ ,
\ee
where $C$ is a constant. The choice of $C$ will fix the total electric
charge of the star. Note that Eq.~(\ref{e:conductparfait}) holds inside
the star only.

\subsection{Global quantities}

The magnetic dipole moment $\cal M$ is given by the leading term of the
asymptotic behaviour of the magnetic field as measured by ${\cal O}_0$:
\be
    B_{(r)} \sim_{r\rightarrow\infty} {\mu_0 \ov 2\pi}
	{{\cal M} \cos\theta \ov r^3}  \ ; \quad
    B_{(\theta)} \sim_{r\rightarrow\infty} {\mu_0 \ov 4\pi}
	{{\cal M} \sin\theta \ov r^3}  \ ,
\ee
$B_{(r)}$ and $B_{(\theta)}$ being the component of $\vec{B}$ in the
orthonormal basis associated to $(r,\theta,\phi)$. They are related to the
components $B_r$ and $B_\theta$ given in Eq.~(\ref{e:B_alpha}) by
\be
    B_{(r)} = {B\ov A^2} \, B_r	 \ ; \quad
    B_{(\theta)} = {B\ov A^2 r} \, B_\theta \ .
\ee

The total electric charge of the star, $Q$,  is given by the leading term
of the asymptotic behaviour of the electric field as measured by ${\cal O}_0$:
\be \label{e:Echarge}
   E_{(r)} \sim_{r\rightarrow\infty} {\mu_0 \ov 4\pi} {Q\ov r^2} \ ; \quad
   E_{(\theta)} \sim_{r\rightarrow\infty} {\mu_0 \ov 4\pi} {Q\ov r^2} \ .
\ee
Equivalently, $Q$ is expressible as
\begin{eqnarray}
   Q & = & \int_{\Sigma_t} j^\mu \, \epsilon_{\mu\alpha\beta\gamma}
		\  + \  Q_{\rm surf} \\
     & = & \int {N A^6\ov B} \, j^t\, r^2 \sin\theta \, dr\, d\theta\, d\phi
		\  + \  Q_{\rm surf} \ ,
\end{eqnarray}
where $Q_{\rm surf}$ is the surface charge.

\section{The numerical procedure and its tests} \label{s:tests}

We describe here only the electromagnetic part of the code. For the fluid and
gravitational part, we refer to BGSM. Let us simply recall that our code is
based on an iterative procedure, each step consisting in solving Poisson
equations by means of a Chebyshev-Legendre spectral method developed by
Bonazzola \& Marck (1990). Typically the computations have been performed
with 41 Chebyshev coefficients in $r$ and 21 coefficients in $\theta$.

\subsection{Numerical resolution of the electromagnetic equations}
\label{s:resol}

In order to obtain a solution of the equations presented in
Sect.~\ref{s:electromagn}, one needs to specify a value for the
central log-enthalpy $H$, a value for the angular velocity $\Omega$, and a
value for the total electric charge $Q$ (possibly $Q=0$).
(An alternative scheme could be, instead of fixing $Q$, to impose
$\vec{E}\cdot\vec{B}=0$ as a boundary condition at the surface of the star in
order to mimic a magnetosphere; the solution would be then an approximate one
outside the star, which is not the case in the procedure described below).
One needs also to pick a choice for the current function $f$.

The iterative procedure is then as follows. At the first step,
$A_\phi$ is set identically to zero. At a given step,
a value ${}^0\! A_t$ of $A_t$ inside the star is deduced from the
value of $A_\phi$ at the previous step by the perfect conductivity equation
(\ref{e:conductparfait}), setting the constant $C$ to zero. Then one
computes the Laplacian of ${}^0\! A_t$ and use the Maxwell-Gauss equation
(\ref{e:MaxGau}) to obtain the value of the charge density $j^t$ inside
the star (using for the $j^\phi$ which appears in Eq.~(\ref{e:MaxGau}) its
value at the previous step). The current density component $j^\phi$ is
then deduced from $j^t$ and the previous value of $A_\phi$ via the
function $f$ and Eq.~(\ref{e:jphi=f(Aphi)}). Then one solves the
Maxwell equations (\ref{e:MaxGau}) and (\ref{e:MaxAmp}) by considering them
as two {\em linear} Poisson equations for $A_t$ and $\tilde A_\phi$,
with a fixed right-hand side: this latter is calculated from the newly
determined $j^t$ and $j^\phi$ and the previous step values of $A_t$ and
$A_\phi$ (in the non-linear terms containing
$\partial A_t$ and $\partial A_\phi$). For $\tilde A_\phi$,
the boundary condition\footnote{as explained in BGSM, our
numerical grid extends to $r=+\infty$ thank to a change of coordinate
$u=1/r$ in the space outside the star.}
$A_\phi=0$ at $r=+\infty$ is used. We then obtain a unique smooth solution
for $A_\phi$. For $A_t$ the procedure is a little more complicated since in
general a rotating perfect conductor is endowed with a surface charge
density, so that the component of the electric field normal to the
surface is discontinuous, meaning that the electric potential $A_t$ is
not differentiable accross the surface. We then solve the equation for $A_t$
in two steps: (i) a solution ${}^1\! A_t$ to the Poisson equation
(\ref{e:MaxGau}) is obtained outside the star with the boundary condition
${}^1\! A_t=0$ at $r=+\infty$. This solution, is by no means unique,
since any harmonic function $\psi$ (i.e. satisfying $\Delta_3 \psi=0$)
vanishing at infinity can be added to ${}^1\! A_t$. (ii) an harmonic function
of the form
\be \label{e:psi,harm}
    \psi(r,\theta) = \sum_{l=0}^L a_l {P_l(\cos\theta) \ov r^{l+1}} \ ,
\ee
where $P_l$ is the Legendre polynomial of degree $l$,
is added to ${}^1\! A_t$ ouside the star so that ${}^0\! A_t := {}^1\! A_t+
\psi$
matches, on the star's surface, the value of ${}^0\! A_t$ inside the star which
has been determined previously by the perfect conductor relation
${}^0\! A_t = -\Omega A_\phi$. Technically the matching is achieved by solving
a linear system for the $L+1$ coefficients $a_l$ of the expansion
(\ref{e:psi,harm}), $L+1$ being the number of grid points on the
star's surface (number of collocation points in $\theta$).
The solution ${}^0\! A_t$ obtained in this way is continuous in all space but
not
necessarily differentiable accross the star's surface. Besides, this solution
has a certain electric charge, ${}^0\! Q$ say, given by
Eq.~(\ref{e:Echarge}) and which does not a priori coincide
with the desired electric charge $Q$. The arbitrary
nature of the constant $C$ of Eq.~(\ref{e:conductparfait}) can be used
to adjust the electric charge, ${}^0\! A_t$ having been obtained by setting
$C=0$. This is achieved by considering the following solution of the
equation $\Delta_3 \, {}^2\! A_t = 0$: ${}^2\! A_t=1$ inside the star and
${}^2\! A_t$ has the same form as the function $\psi$ of Eq.~(\ref{e:psi,harm})
outside the star,
the coefficients $a_l$ being determined in order to insure the continuity
of ${}^2\! A_t$ accross the star's surface. Let ${}^2\!Q$ be the electric
charge
associated with ${}^2\! A_t$. The constant $C$ is then determined by
\be
    C = {Q - {}^0\! Q \ov {}^2\! Q}
\ee
and the final solution for $A_t$ is
\begin{eqnarray}
    A_t & = & {}^0\! A_t + C = -\Omega A_\phi + C \quad \mbox{inside the star}
\\
    A_t & = & {}^0\! A_t + C\  {}^2\! A_t \quad \mbox{outside the star} \ .
\end{eqnarray}
By construction $A_t$ satisfies the Poisson equation (\ref{e:MaxGau}) (with
the right-hand computed from values at the previous step) and
corresponds to the required electric charge $Q$.

Once $A_t$ and $A_\phi$ are obtained, the electric and magnetic fields are
computed via Eqs.~(\ref{e:E_alpha}) and (\ref{e:B_alpha}).
The electromagnetic stress-energy tensor is then deduced from Eq.~(5.32)
of BGSM and put as a source term, beside the fluid stress-energy
tensor, in the Einstein equations giving the gravitational field
(Eqs.~(3.19)-(3.22) of BGSM).

\subsection{Tests of the numerical code}

Various tests have been performed on the electromagnetic part of the code
(for the gravitational part we refer to BGSM), mainly by comparing
numerical solutions to exact ones for simple charge and current
distributions.

\subsubsection{Rotating magnetic dipole} \label{s:rot,mag,dip}

We have considered an electrostatic dipole, a magnetostatic dipole
and a rotating magnetic dipole, all of them in flat space (no gravitation).
Let us detail the results for the rotating magnetic dipole, since they are
representative of the precision achieved in all other cases.

By ``rotating magnetic dipole'', we mean two different physical situations:
(i) a conducting sphere in rotation with surface electric currents and
(ii) a conducting sphere in rotation with an infinitely small electric current
loop at the origin. In case (i), the current distribution is modelled by
\be
    j^{(\phi)}(r,\theta) = j_0 \, r \sin\theta \exp\l(
	-{ (r-R)^2 \over \delta^2 } \r) \ ,
\ee
where $j_0$ is a constant, $R$ is the sphere's radius and $\delta$ is the
width of the distribution, which we do not put to zero in order to have
only smooth functions. Typically, we use $\delta \sim .01\, R$.
The exact solution for $(\vec{E},\vec{B})$ (corresponding to $\delta=0$)
may be found in Ruffini \& Treves (1973). The magnetic field is
uniform and parallel to the rotation axis inside the sphere and dipolar
outside.
The electric field outside the sphere is the sum of a monopolar term and a
quadrupolar one.

\begin{figure}
\psfig{figure=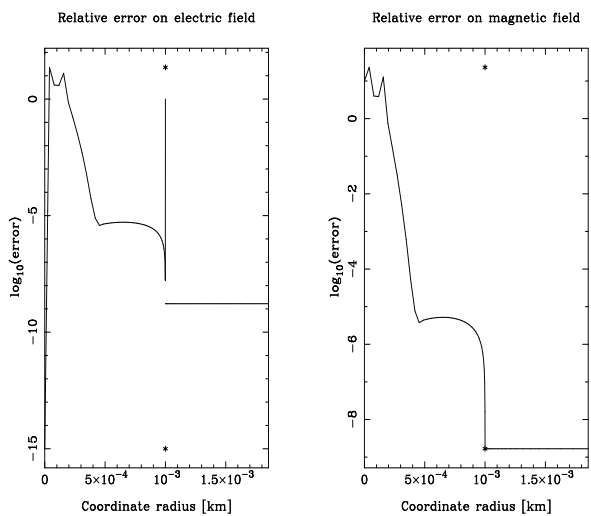,angle=270,height=7cm,width=8.5cm}
\caption[]{\label{f:test}
Comparison between the exact solution and the numerical one in the case of a
rotating conducting sphere with a point magnetic dipole moment at its centre.
The left figure corresponds to the electric field, the error being defined
as the square root of $[ ( E_{(r)} - E_{(r)}^{\rm exact} )^2
+ ( E_{(\theta)} - E_{(\theta)}^{\rm exact} )^2 ] / [  (E_{(r)})^2 +
(E_{(\theta)})^2 ] $. The right figure depicts the same thing for the
magnetic field. Asterisks denote the sphere's surface.}
\end{figure}

In case (ii), the current distribution is modelled by
\be
    j^{(\phi)}(r,\theta) = j_0 \, r \sin\theta \exp\l(
	-{ r^2 \over \delta^2 } \r) \ .
\ee
The exact solution (corresponding to $\delta=0$, i.e. to a point magnetic
dipole moment) may be found in the Table~4.1 of Michel (1991)\footnote{in this
reference, the term $2/3r$ in the seventh line should read $2/3r^2$}.
The magnetic field is dipolar in all space. The electric field outside the
sphere is the sum of a monopolar term and a quadrupolar one.
The difference between the exact solution and the solution given by the
numerical code is depicted in Fig.~\ref{f:test} in the case of
a conducting sphere of diameter 1 m, rotating at the angular velocity
$\Omega=3000{\rm \ rad\, s}^{-1}$, with a central loop current
$j_0 = 10^{11} {\rm \ A\, m}^{-2}$.
Near the star's centre, the discrepancy is quite important because the current
distribution used in the code is not exactly a point dipole ($\delta\not = 0$).
But at half the stellar radius, the relative error becomes less than $10^{-5}$
and is of the order $\sim 10^{-9}$ outside the star.

\subsubsection{Comparison with Ferraro's solution}

Ferraro (1954) has obtained an analytical solution for a Newtonian
incompressible fluid body with an electric current distribution
corresponding to the choice $f(x)={\rm const}$ for the current
function of Eq.~(\ref{e:jphi=f(Aphi)}). Ferraro's solution is valid for a
body whose shape remains close to a sphere and reads
\begin{eqnarray}
  A_\phi & = & - \mu_0 \rho c^2 f_0 ( r^2 / 10 - R^2/6 )
         r^2 \sin^2\theta
         \quad \mbox{for\ } r\leq R \\
  A_\phi & = & \mu_0 \rho c^2  f_0 {R^5\ov 15 r} \sin^2 \theta
         \qquad \mbox{for\ } r\geq R \ ,
\end{eqnarray}
where $f_0$ is the constant value taken by the current function $f$ and
$R$ is the stellar radius. The discrepancy between Ferraro's solution
and the numerical one resulting from the code is depicted in
Fig.~\ref{f:test,Ferraro}. It is everywhere less than $10^{-3}$, being
less than $5\times 10^{-4}$ inside the star. These numbers
are higher than those of the test of Sect.~\ref{s:rot,mag,dip} but one
should keep in mind that Ferraro's solution is only an approximate one,
assuming no deviation from spherical symmetry, whereas the numerical
solution takes into account the deformation of the star by Lorentz
forces. Of course, in order to perform the test, we have considered
a magnetic field weak enough for Ferraro's approximation
to be valid, as well as a very weak gravitational field for the Newtonian
approximation to be justified: the tested configuration is a
$2.67\times 10^{-2} \, M_\odot$ star of (constant) density $1.66\times
10^9\ {\rm kg\, m}^{-3}$ with a polar magnetic field $B_{\rm pole} = 6.6\times
10^5$~T.

Ferraro's solution is interesting since it allows to test the
response of the star to the magnetic field, contrary to the tests of
Sect.~\ref{s:rot,mag,dip}.
The deformation of the star as computed by
Ferraro at first order around the spherical symmetry, is given in terms
of the polar magnetic field by the eccentricity
\be
    \epsilon = {15\ov 4} {B_{\rm pole} \ov \sqrt{\pi G \mu_0} \, \rho R} \ .
\ee
For the configuration considered above, $\epsilon=0.04670$. The eccentricity
resulting from the code is $0.04683$, so that the relative discrepancy is
$3\times10^{-3}$. Here again let us recall that Ferraro's solution is
only an approximate one, so that this value reflects not only the true
numerical
error but also Ferraro's error with respect to the exact solution.
Nevertheless the very good agreement with Ferraro's solution means
that the action of Lorentz forces upon the fluid is correctly
treated by the code and in particular that there is no error in the
conversion from magnetic units to fluid units.

\begin{figure}
\psfig{figure=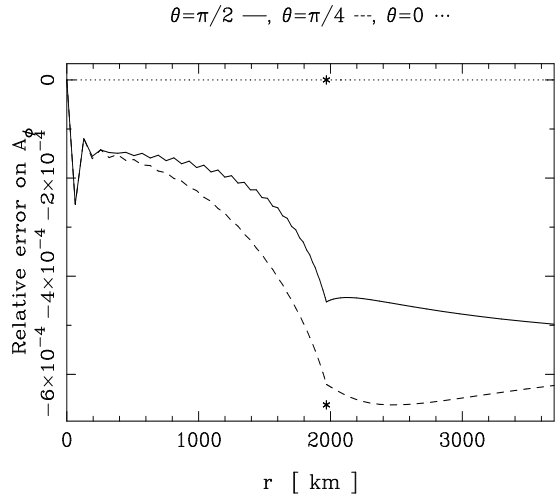,angle=270,height=7cm,width=8.5cm}
\caption[]{\label{f:test,Ferraro}
Comparison between Ferraro's analytical solution and the numerical one in the
case of a Newtonian incompressible fluid endowed with a magnetic field
corresponding to the current function $f(x) = {\rm const}$.
The plotted quantity is the relative difference between the two values
of $A_\phi$ as a function of the radial coordinate $r$ for three values
of $\theta$. Asterisks denote the star's surface.}
\end{figure}

\subsubsection{Virial identities}

A different kind of test is provided by the virial identities GRV2
(Bonazzola \& Gourgoulhon 1994) and GRV3 (Gourgoulhon \& Bonazzola 1994),
the latter being a relativistic generalization of the Newtonian virial
theorem.
GRV2 leads to the $|1-\lambda|$ error indicator introduced in BGSM.
These virial identities allow to test each computation and not only the
simplified ones presented above. It notably controls the convergence of
the iterative procedure used in the code. In most of our calculations, the
final value of the GRV2 or GRV3 error indicator is of the order
$10^{-5}$ (polytropic EOS) or a few $10^{-4}$ (tabulated EOS). Let us recall
that tabulated
EOS introduce an important numerical error because they do not stricly
obey to the thermodynamical relations (cf. Sect.~4.2 of Salgado et al. 1994).
We systematically
rejected any solution for which the GRV2 or GRV3 error is greater than
$10^{-2}$, such a relative
error being too large for considering that the numerical procedure has
converged. This may happen if the magnetic field is too large for any
stationary configuration to exist.

\begin{figure}
\psfig{figure=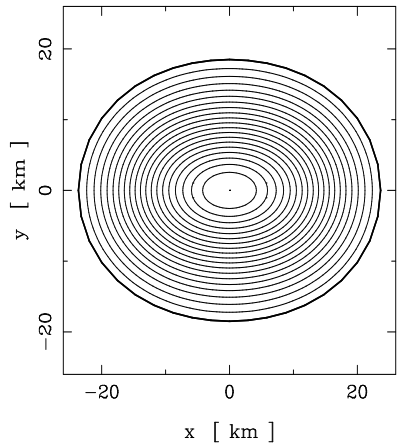,angle=270,height=7cm,width=8.5cm}
\caption[]{\label{f:static,isoener,4E4}
Fluid proper density isocontours in the $(r,\theta)$ plane of a static
magnetized star of $2.98\, M_\odot$
built on the Pol2 EOS of Salgado et al. (1994a) (polytropic EOS with
$\gamma=2$) with a current function $f(x)={\rm const.} =
4\times 10^{15}/R {\ \rm A\, m}^{-2} \rho_{\rm nuc}^{-1} {\rm c}^{-2}$,
where $R$ is the $r$ coordinate of the star's equator. The central energy
density
is $e=1.42\, \rho_{\rm nuc} {\rm c}^2$
($1 \, \rho_{\rm nuc}:=1.66\times 10^{17}\ {\rm kg\, m}^{-3}$).}

\end{figure}

\begin{figure}
\psfig{figure=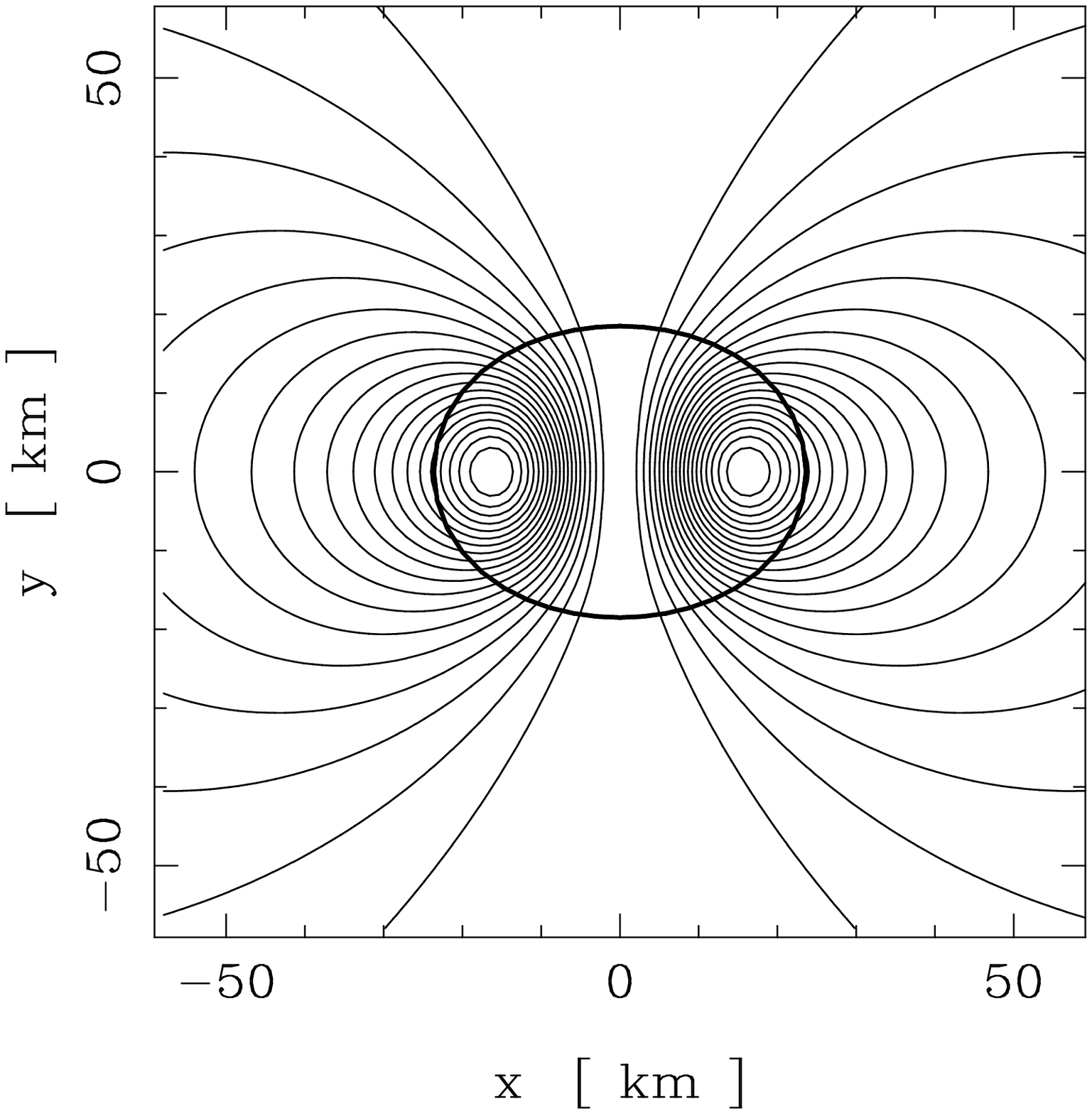,angle=270,height=7cm,width=8.5cm}
\caption[]{\label{f:static,B,4E4}
Magnetic field lines in the $(r,\theta)$ plane for the static star
configuration corresponding to
Fig.~\ref{f:static,isoener,4E4}. The thick line denotes the star's surface.
The magnetic field amplitude is $B_{\rm c} = 3.57\times 10^4 {\ \rm GT}$ at the
star's centre and $B_{\rm pole} = 9.1\times 10^3 {\ \rm GT}$ at the north
pole; the magnetic dipole moment is
${\cal M} = 5.42\times 10^{32} {\ \rm A\, m}^2$.}
\end{figure}

\begin{figure}
\psfig{figure=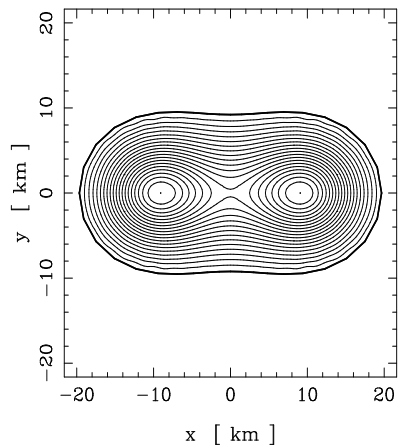,angle=270,height=7cm,width=8.5cm}
\caption[]{\label{f:static,isoener,Mmax}
Fluid proper density isocontours in the $(r,\theta)$ plane
for the maximum mass static magnetized star
built on the Pol2 EOS of Salgado et al. (1994a) (polytropic EOS with
$\gamma=2$). The current function $f(x)={\rm const.} = 5.63\times 10^{15}
 /R {\ \rm A\, m}^{-2} \rho_{\rm nuc}^{-1} {\rm c}^{-2}$,
where $R$ is the $r$ coordinate of the star's equator.
The central energy density is $e=1.42\, \rho_{\rm nuc} {\rm c}^2$ and the
mass $M=4.06\, M_\odot$; the other characteristics are listed on the third
line of Table~\ref{t:static,Mmax}.}
\end{figure}

\begin{figure}
\psfig{figure=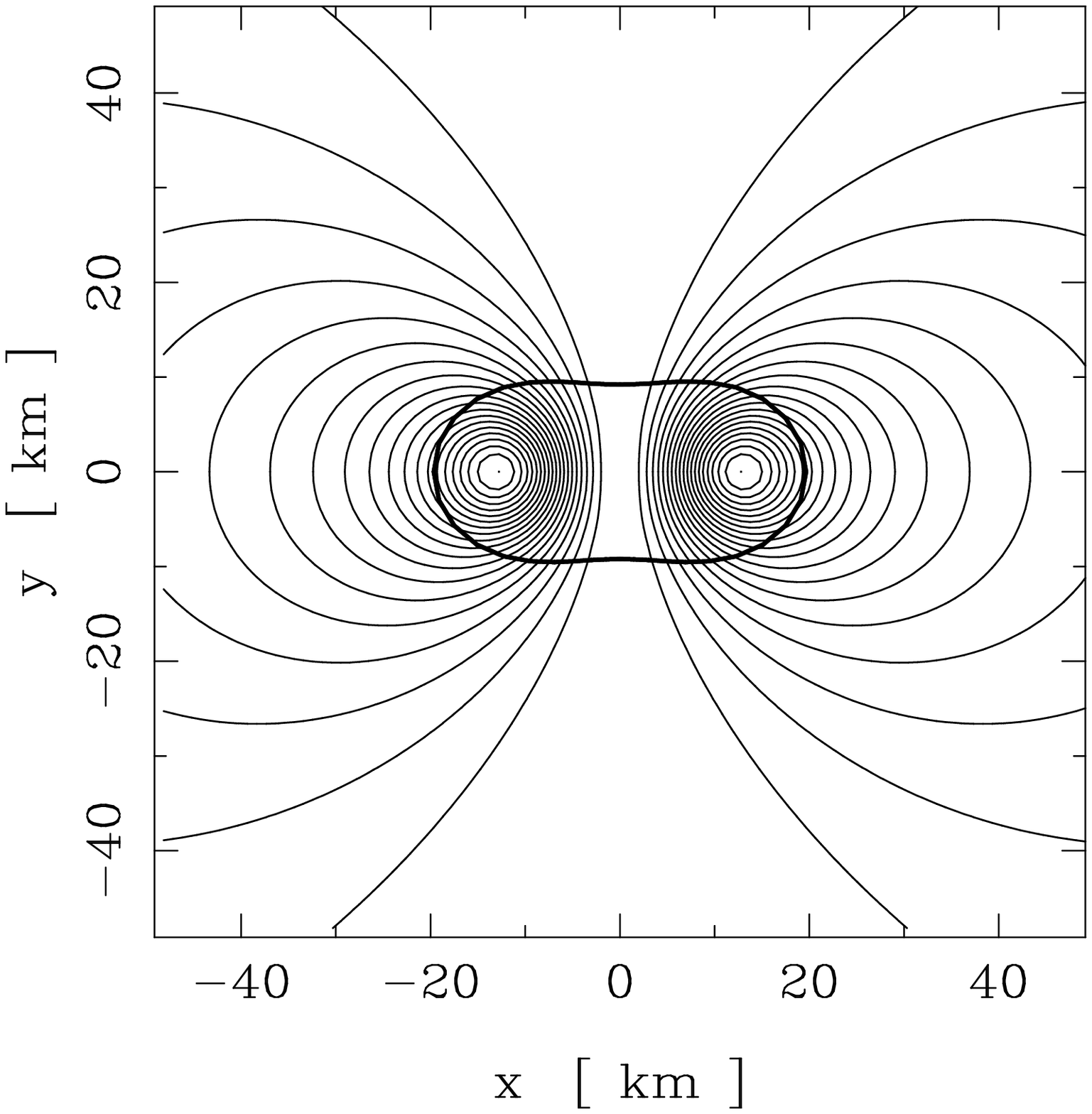,angle=270,height=7cm,width=8.5cm}
\caption[]{\label{f:static,B,Mmax}
Magnetic field lines in the $(r,\theta)$ plane for the maximum mass
configuration corresponding to
Fig.~\ref{f:static,isoener,Mmax}. The thick line denotes the star's surface.
The magnetic field amplitude is $B_{\rm c} = 9.00\times 10^4 {\ \rm GT}$ at the
star's centre and $B_{\rm pole} = 4.57\times 10^4 {\ \rm GT}$ at the north
pole; the magnetic dipole moment is
${\cal M} = 1.12\times 10^{33} {\ \rm A\, m}^2$.}
\end{figure}

\begin{figure}
\psfig{figure=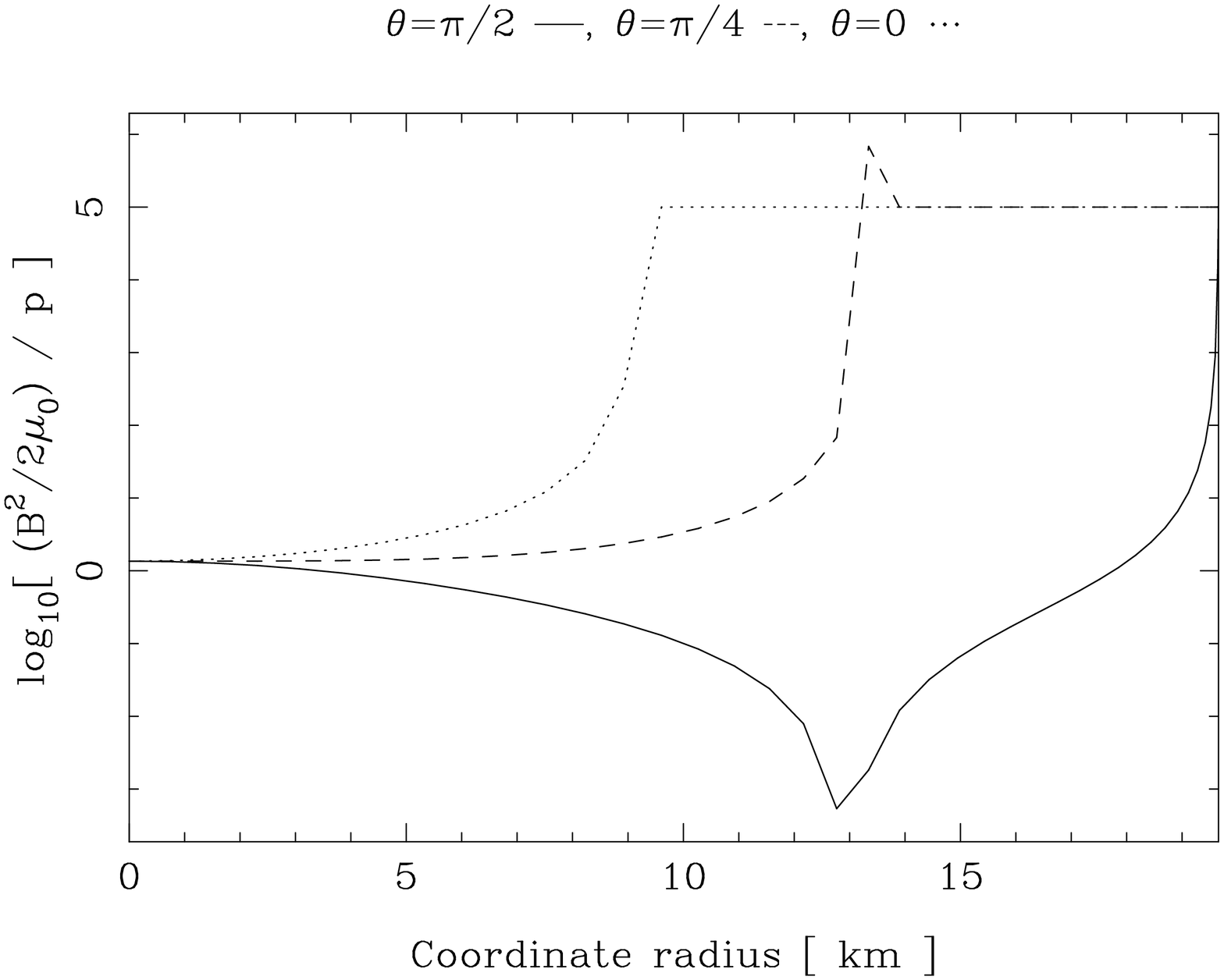,angle=270,height=7cm,width=8.5cm}
\caption[]{\label{f:static,rappress}
Ratio of the magnetic pressure $B^2/2\mu_0$ to the fluid pressure $p$ as
a function of the radial coordinate $r$ and along three angular directions
for the maximum mass configuration displayed in
Figs.~\ref{f:static,isoener,Mmax} and \ref{f:static,B,Mmax}.
Outside the star, where $p=0$, the logarithm of the ratio has been set to 5.
At the star's centre, $(B^2/2\mu_0)/p = 1.36$.}
\end{figure}

\section{Static magnetized configurations} \label{s:static}

In order to investigate some purely magnetic effects on stellar configurations,
let us first consider the case of static (i.e. non rotating) neutron
stars. More precisely, a {\em static} spacetime (in the sense of the time
Killing vector being hypersurface-orthogonal) implies not only a non rotating
fluid ($\Omega=0$) but also a vanishing electric charge. Only in this
case the electric field as measured by ${\cal O}_0$ is zero. On the contrary,
a non-vanishing electric charge creates an electrostatic field outside the
star,
so that there exists a net
electromagnetic momentum density vector $\vec{E}\times\vec{B}/2\mu_0$
which generates a non-zero angular momentum, though the fluid does not rotate
[Feynman's disk paradox (Feynman et al. 1964, Ma 1986, de Castro 1991)].
Consequently the shift vector component
$N^\phi \not = 0$ and the spacetime is not static.
We observed this effect numerically but only for huge values of the electric
charge: $Q \sim 10^{17}{\ \rm C}$.

\subsection{Deformation of the star} \label{s:static:deform}

Even in the static case, the stress-energy tensor of the electromagnetic
field is not isotropic (cf. Eq.~(5.33) of BGSM with $U=0$) so that the
star deviates from spherical symmetry. For large magnetic field this
deviation  becomes important as shown in Fig.~\ref{f:static,isoener,4E4}.
The corresponding
magnetic field lines are
plotted in Fig.~\ref{f:static,B,4E4}. Note that the situation depicted in
Figs.~\ref{f:static,isoener,4E4} and \ref{f:static,B,4E4}
corresponds to an important magnetic field:
$B_{\rm pole} = 9.1\times 10^3 {\ \rm GT}$.
The flattened shape is a general result for a magnetized configuration.
We thus may conclude that the Lorentz forces exerted by the
electromagnetic field on the conducting fluid behave as centrifugal
forces. Note that the flattening property of the magnetic field has
been recognized long ago (in the Newtonian case)
by Chandrasekhar \& Fermi (1953).

\subsection{Maximum values of the magnetic field} \label{s:Bmax}

If the value of the central magnetic field is high enough for the magnetic
pressure equals the fluid pressure $p$, then the total stress tensor $S_{ij}$
has a vanishing component along the symmetry axis, as it can be seen by
combining Eq.~(5.7c) and (5.32c) of BGSM:
\be
    S_r^{\ \, r} = p + {1\ov 2\mu_0} \l( B_\theta B^\theta - B_r B^r \r) \ ,
\ee
with, on the symmetry axis near the centre,
$B_\theta B^\theta = 0$ and $B_r B^r /2\mu_0 \simeq p$. Hence
$S_r^{\ \, r}=0$. Away from the centre, the fluid pressure decreases more
rapidly than the magnetic pressure along the symmetry axis, so that
$S_r^{\ \, r}<0$, which means that the
fluid + magnetic field medium develops some {\em tension} instead of
{\em pressure} along the symmetry axis. The star displays then a pinch along
the symmetry axis, as it can be seen on Figs.~\ref{f:static,isoener,Mmax}
and \ref{f:static,B,Mmax}. The fact that in this case the magnetic pressure
exceeds the fluid pressure everywhere on the symmetry axis can be clearly seen
in Fig.~\ref{f:static,rappress}
(dotted line), which represents the ratio of the two pressures throughout
the star. For magnetic fields higher than that presented in
Figs.~\ref{f:static,isoener,Mmax}-\ref{f:static,rappress}, no
stationary configuration can exist and the numerical procedure
described in Sect.~\ref{s:resol} fails to converge.

The stellar shape displayed in Fig.~\ref{f:static,isoener,Mmax} is
a typical feature of the maximum allowable magnetic field and is not due to
the fact that this particular configuration is the maximum mass one or
that the EOS is Pol2.
Indeed, we found very similar shapes for all central densities in the
explored range ($0.15 \rho_{\rm nuc}$ up to $5 \rho_{\rm nuc}$) and
all EOS when the
magnetic field has reached the maximum value for which a solution could be
obtained. In all cases, the ratio of $B^2/2\mu_0$ to $p$ was close
to 1 at the star's centre (cf. the sixth column of Table~\ref{t:static,Mmax}).

Note that the maximum values of $B_{\rm pole}$ are four orders of magnitude
higher than the maximum observed value: $B_{\rm pole} = 2.1 {\ \rm GT}$ in
the pulsar PSR B0154+61 (Taylor et al. 1993), which means that the
magnetic field in the observed pulsar is not limited by the condition of
stellar equilibrium but by the physical processes it originates from.

\subsection{Effect of the magnetic field on $M_{\rm max}$}
\label{s:static,effect/Mmax}

In order to investigate the effect of the magnetic field on the maximum
mass of static configurations, we choose the simplest form of the current
function:
\be
   f(x) = {\rm const.} = f_0 \ ,		\label{e:f(x)=const}
\ee
and vary $f_0$ from $0$ (no magnetic field) up to the maximal
value for which the code converges and for which the ratio of the
magnetic pressure to the fluid pressure at the star's centre come close to one
(cf. sixth column of Table~\ref{t:static,Mmax}).
The choice (\ref{e:f(x)=const}) results in electric currents concentrated
deep inside the star, with the same behaviour as that represented in
Fig.~\ref{f:jphi,f=const} for the rotating case.
Each static configuration is determined by
the value of the central log-enthalpy $H_{\rm c}$ and the value of $f_0$.
In the plane $(H_{\rm c},f_0)$, we followed curves of constant magnetic
dipole moment $\cal M$ and determined for each of them the maximal
gravitational mass.

This study has been conducted for five of the EOS used in Salgado et al.
(1994a,b): Pol2 ($\gamma=2$ polytrope), DiazII (model II of Diaz Alonso 1985),
PandN (pure neutron model of Pandharipande 1971), BJI (model IH of
Bethe \& Johnson 1974) and HKP (Haensel et al. 1981).
For details about these EOS, the reader may consult Sect.~4.1 of Salgado
et al. (1994a).

Some results are presented in Table~\ref{t:static,Mmax} which gives
maximum mass configurations (i) with no magnetic field (first line),
(ii) with some fixed magnetic dipole moment (second line), (iii)
among all magnetized configurations (third line).

Different behaviours appear for different EOS:
the maximum gravitational mass at fixed
magnetic dipole moment $\cal M$ as well as the maximum
baryon number at fixed $\cal M$ are both increasing functions of $\cal M$
for the EOS DiazII (cf. Fig.~\ref{f:static,Mmax,Diaz}), whereas the maximum
baryon number is a decreasing function of $\cal M$ for the EOS BJI
(cf. Fig.~\ref{f:static,Mmax,BJI}), the maximum mass remaining a increasing
function. Pol2 models behave as DiazII ones, whereas PandN and HKP models
behave as BJI ones.
In all cases, the maximum gravitational mass increases with the magnetic
field, by $28.6\%$ for Pol2,
$20.2\%$ for DiazII, $14.9\%$ for PandN, $17.3\%$ for BJI,  and
$13.3\%$ for HKP (compare first and third lines in Table~\ref{t:static,Mmax}).

\begin{table*}
\caption[]{\label{t:static,Mmax}
Maximum mass (at fixed magnetic dipole moment) neutron stars
in the non-rotating case. For each EOS, the last line is the maximum mass
configuration among all static magnetized models.
$\cal M$ is the magnetic dipole moment, $B_{\rm c}$ and $B_{\rm pole}$ are
respectively the values of the magnetic field at the centre and the north
pole of the star,
$H_{\rm c}$ is the central value of the log-enthalpy as defined
by Eq.~(\ref{e:def:H}),
$e_{\rm c}$ the central energy density
($1\, \rho_{\rm nuc} := 1.66\ 10^{17}\, {\rm kg\, m}^{-3}$),
$p_{\rm mg,c} / p_{\rm fl,c}$ the ratio of the magnetic pressure to the
fluid pressure at the star centre,
$M$ the gravitational mass, ${\cal B}$ the baryon mass,
$E_{\rm bind}$ the binding energy per baryon (with the convention
$E_{\rm bind}<0$ for a bound configuration),
 $R_{\rm circ}$ the
circumferential radius (length of the star's equator divided by $2\pi$) and
GRV2 and GRV3 the estimates of the global numerical relative error
provided by the virial identities GRV2 and GRV3 (cf. text). We use
$G=6.6726\ 10^{-11}\ {\rm m}^3{\rm kg}^{-1}{\rm s}^{-2}$ and
$1\ M_\odot=1.989\ 10^{30}\ {\rm kg}$.}
\begin{flushleft}
\begin{tabular}{rrrrrrrrrrrrr}
\hline\noalign{\smallskip}
 $\displaystyle {{\rm EOS} \atop \ }$  &
 $\displaystyle {{\cal M} \, [10^{32} \atop  {\rm A\, m}^2]}$  &
 $\displaystyle {B_{\rm c} \atop [10^{3}\, {\rm  GT}]}$ &
 $\displaystyle {B_{\rm pole} \atop [10^{3}\, {\rm  GT}]}$ &
 $\displaystyle {H_{\rm c} \atop \ }$ &
 $\displaystyle {e_{\rm c}\atop [\rho_{\rm nuc} c^2]}$  &
 $\displaystyle {p_{\rm mg,c} \ov p_{\rm fl,c}}$  &
 $\displaystyle {M\atop [M_\odot]}$ &
 $\displaystyle {{\cal B}\atop [M_\odot]}$  &
 $\displaystyle {E_{\rm bind}\atop [m_{\rm B} c^2]}$  &
 $\displaystyle {R_{\rm circ}\atop [{\rm km}]}$  &
 $\displaystyle {{\rm GRV2}\atop \ }$  &
 $\displaystyle {{\rm GRV3}\atop \ }$  \\
\noalign{\smallskip}
\hline\noalign{\smallskip}
Pol2   &0&0&0&    0.491& 4.17&0&3.158& 3.470&-0.0897&21.77& 1E-14& 4E-12\\
     &2.00&37.8& 6.2&   0.483&4.07&0.04&3.182& 3.486&-0.0872&21.93& 1E-06&
2E-06\\
    &11.22&90.0& 45.7&   0.225&1.42&1.36&4.062& 4.279&-0.0507&26.45& 1E-03&
7E-04\\
 \noalign{\smallskip}
\hline\noalign{\smallskip}
DiazII   &0&0&0&   0.610&15.21&0&1.933&2.210&-0.1253&10.92&1E-04&1E-04\\
     &0.50&73.8&14.1&  0.600&14.85&0.03&1.940&2.212&-0.1229&11.00&2E-04&2E-04\\
    &3.08&207.4&108.3&  0.285&6.22&1.10&2.324&2.508&-0.0734&13.12&1E-04&9E-05\\
 \noalign{\smallskip}
\hline\noalign{\smallskip}
PandN  &0&0&0&   0.733&24.39&0&1.662&1.932&-0.1340&8.55& 1E-04&2E-04\\
       &0.20&64.7&11.9&   0.727&24.13&0.01&1.663&1.931&-0.1384&8.57&
1E-04&2E-04\\
      &1.86&302.8&153.8&   0.350&11.21&1.01&1.910&2.064&-0.0746&10.00&
3E-04&2E-04\\
 \noalign{\smallskip}
\hline\noalign{\smallskip}
BJI     &0&0&0&    0.699&18.64&0&1.856&2.134&-0.1303&9.91& 2E-06&3E-06\\
      &0.30&61.5&11.0&   0.692&18.40&0.01&1.858&2.132&-0.1288&9.94&
2E-06&2E-06\\
     &2.63&233.2&121.0&   0.300&7.47&1.10&2.176&2.344&-0.0717&12.05&
1E-04&5E-05\\
 \noalign{\smallskip}
\hline\noalign{\smallskip}
HKP    &0&0&0&   0.725&8.75&0&2.836&3.422&-0.1712&13.67&1E-04&8E-05\\
      &0.80&56.0&12.4&  0.714&8.57&0.02&2.840&3.417&-0.1689&13.75&1E-04&9E-05\\
     &4.90&196.6&104.5&
0.370&4.36&0.90&3.212&3.594&-0.1063&15.85&6E-04&3E-04\\
\noalign{\smallskip}
\hline
\end{tabular}
\end{flushleft}
\end{table*}

\begin{figure*}
\psfig{figure=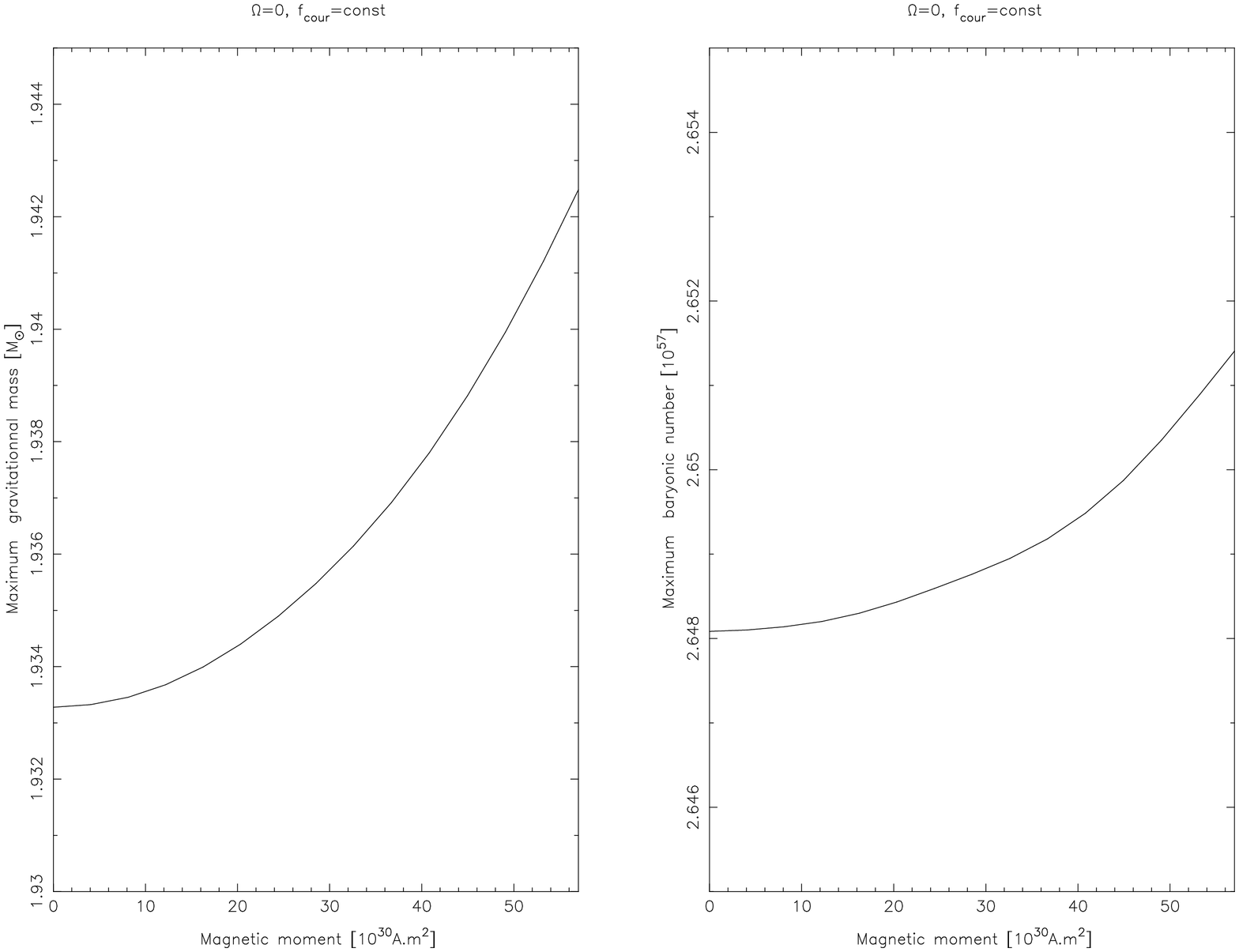,angle=270,height=8cm,width=17.5cm}
\caption[]{\label{f:static,Mmax,Diaz}
Maximum gravitational mass (left) and maximum baryon number
(right) along static sequences at fixed magnetic dipole moment
as a function of the magnetic dipole moment for the DiazII EOS
.}
\end{figure*}

\begin{figure*}
\psfig{figure=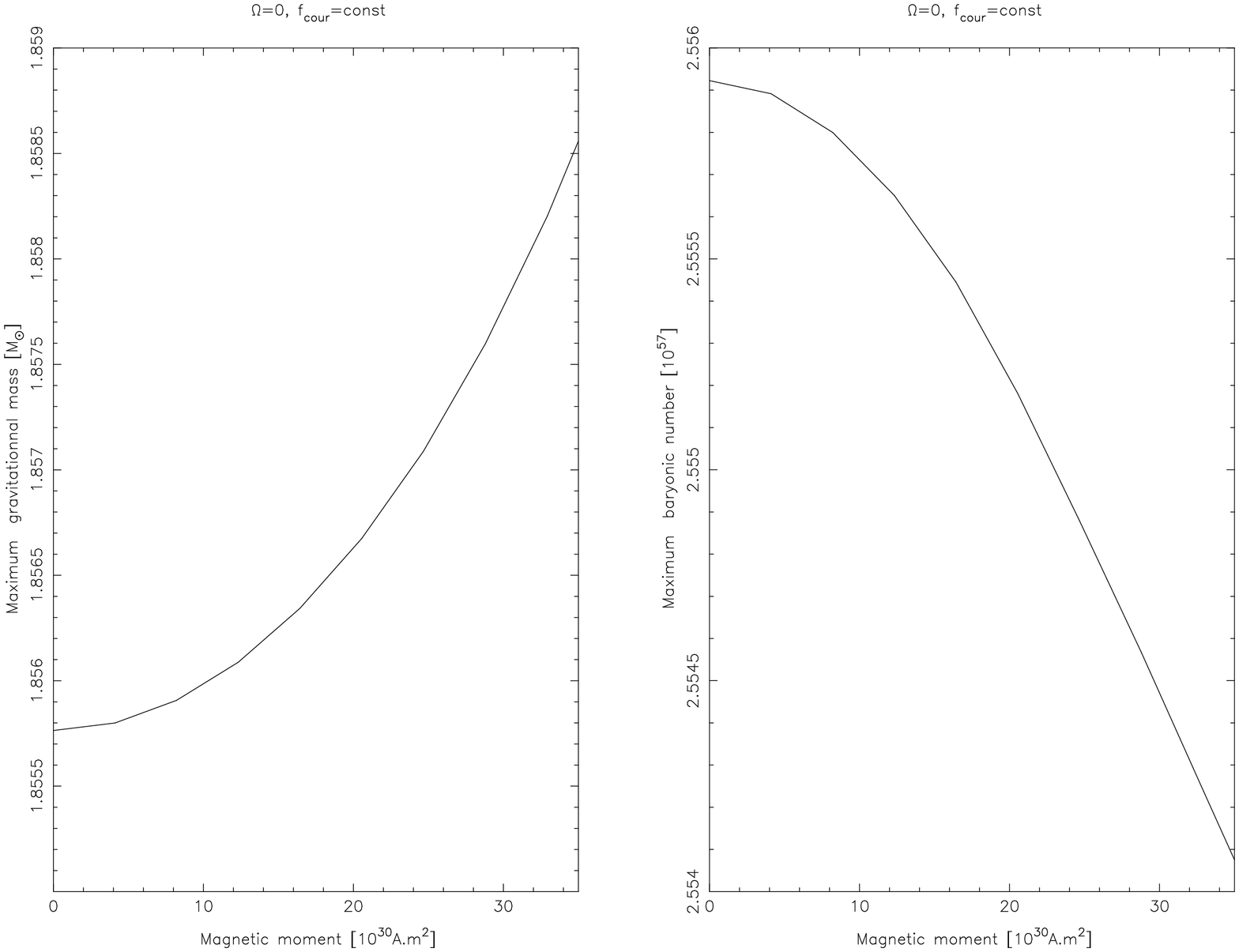,angle=270,height=8cm,width=17.5cm}
\caption[]{\label{f:static,Mmax,BJI}
Same as Fig.~\ref{f:static,Mmax,Diaz} but for the BJI EOS
.}
\end{figure*}

\section{Rotating magnetized configurations} \label{s:rotat}

\subsection{Electric charge}

As we have seen in Sect.~\ref{s:electromagn} and \ref{s:resol}, the total
electric charge $Q$ is a freely specifiable parameter of our models.
Realistic neutron stars certainly possess a net electric charge, which may be
of
the order of $10^{12}$ C (Michel 1991, Chap.~4 and Cohen et al. 1975). The
key point
is that this (huge) charge is compensated by the charge provided by the
magnetospheric particles  to lead to globally neutral star + magnetosphere
system. Now, as stated in Sect.~\ref{s:intro}, we do not attempt to model any
filled magnetosphere and consider neutron stars surrounded by matter vacuum.
In consequence, we do not try to give a specific charge to our models (this
would require a detailed magnetosphere model) and consider globally neutral
stellar configurations: $Q=0$. Anyway, the effect of a non-zero value of $Q$ on
the non-electromagnetic characteristics of the star reveals to be extremely
small:
we verified that
\begin{eqnarray}
   \forall\ H_{\rm c},\Omega, & &
	| M(H_{\rm c},\Omega,Q=10^{18}{\ \rm C}) -  \nonumber \\
  & & \qquad M(H_{\rm c},\Omega,Q=0) | \  \leq \  10^{-5} \ M_\odot \ ,
\end{eqnarray}
which is of the order of the numerical error committed by the code.
The global quantity which is the most sensitive to a non-zero value of $Q$ is
the angular momentum as mentioned in Sect.~\ref{s:static}. In the rotating
case however the fluid angular momentum dominates by at least one order of
magnitude the electromagnetic angular momentum, even for huge values of $Q$
($Q\sim 10^{18}{\ \rm C}$).

\begin{figure}
\psfig{figure=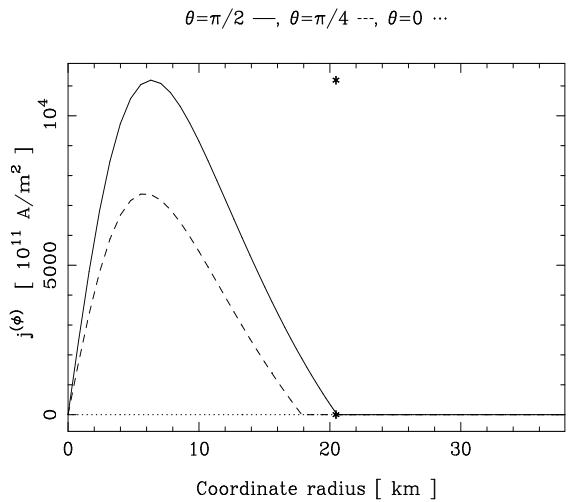,angle=270,height=7cm,width=8.5cm}
\caption[]{\label{f:jphi,f=const}
Electric current distribution induced by the current function
$f(x)= {\rm const.} = 10^{15} /R {\ \rm A\, m}^{-2} \rho_{\rm nuc}^{-1} {\rm
c}^{-2}$
for a Pol2 EOS model,
of central energy density $e_{\rm c}=3.06 \ \rho_{\rm nuc}\, c^2$, rotating
at $\Omega= 3\times 10^3 {\ \rm rad\, s}^{-1}$.
The plotted quantity is  the value
$j^{(\phi)}$ of the azimuthal component of $\vec{j}$ in an orthonormal basis
in the equatorial plane $\theta={\pi/2}$ (solide line), along $\theta={\pi/4}$
(dashed line) and along the rotation axis $\theta=0$ (dotted line).
Asterisks denote the star's equatorial radius.}
\end{figure}

\begin{figure}
\psfig{figure=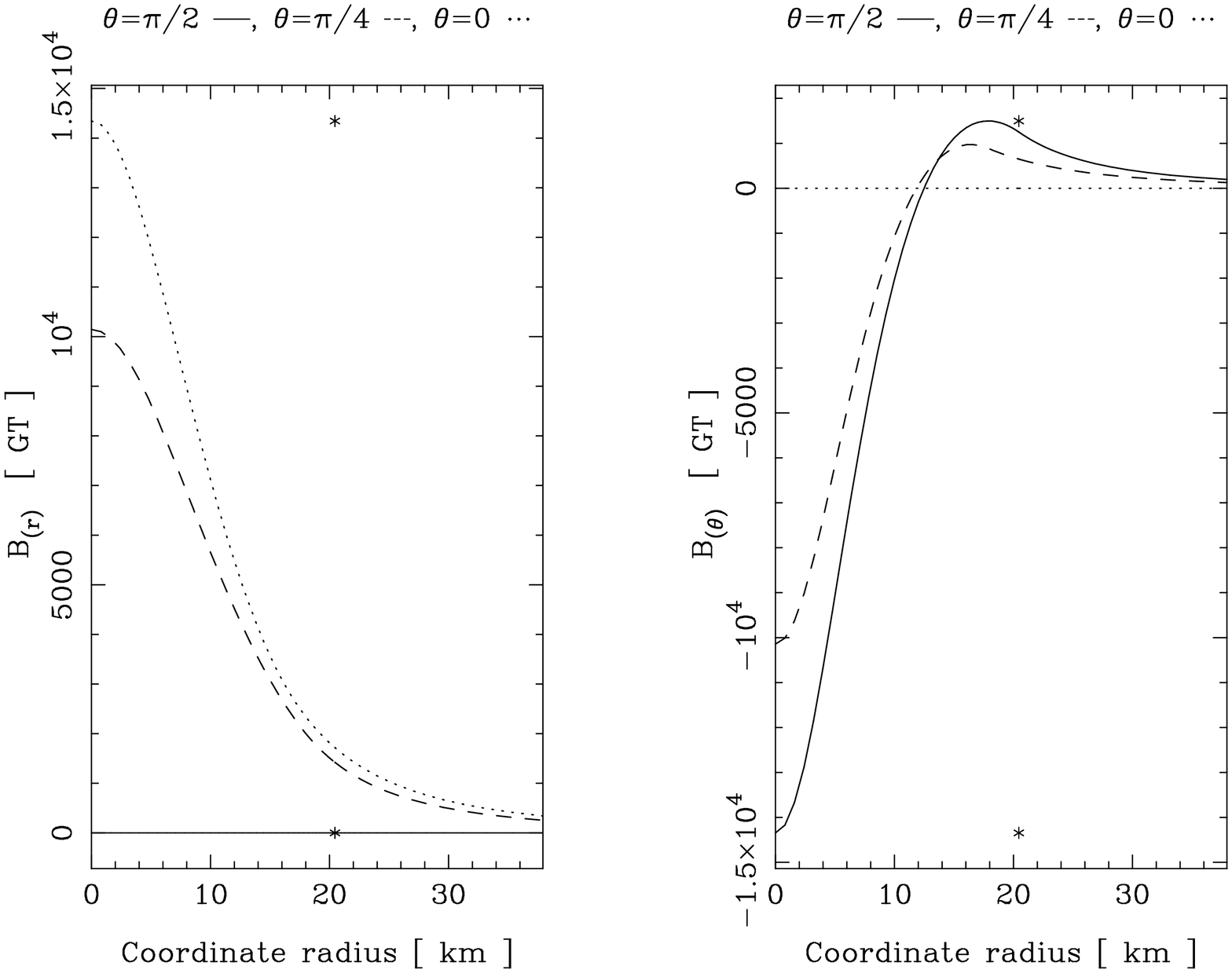,angle=270,height=7cm,width=8.5cm}
\caption[]{\label{f:B,f=const}
Components $B_{(r)}$ (left) and $B_{(\theta)}$ (right) of the magnetic field
generated by the electric current distribution of Fig.~\ref{f:jphi,f=const}.
The notations are the same as in Fig.~\ref{f:jphi,f=const}.}
\end{figure}

\begin{figure}
\psfig{figure=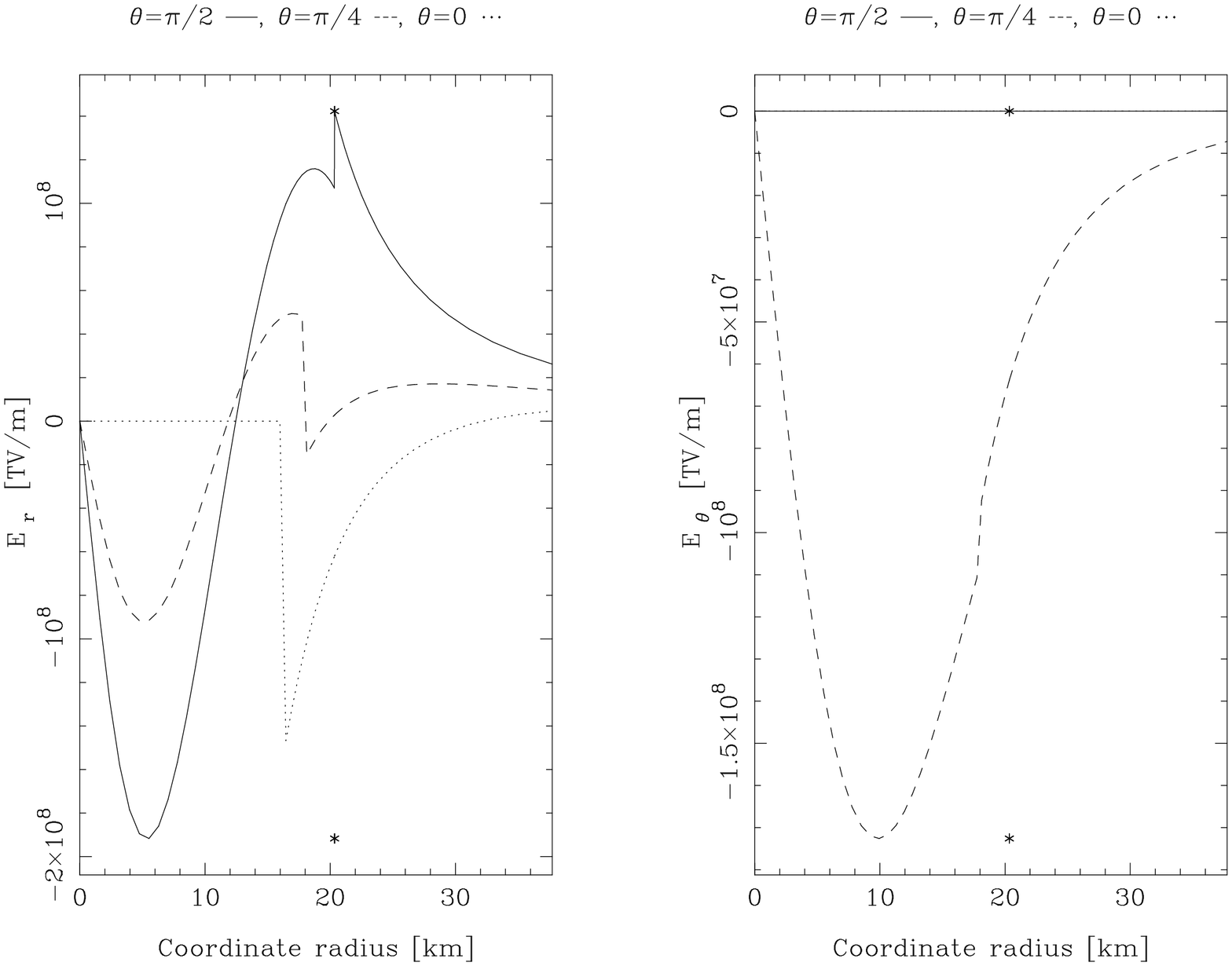,angle=270,height=7cm,width=8.5cm}
\caption[]{\label{f:E,f=const}
Components $E_{(r)}$ (left) and $E_{(\theta)}$ (right) of the electric field
corresponding to the configuration of Figs.~\ref{f:jphi,f=const} and
\ref{f:B,f=const}. The notations are the same as in Fig.~\ref{f:jphi,f=const}.
On the right figure, the dotted line merges with the solid one.}
\end{figure}

\begin{figure*}
\psfig{figure=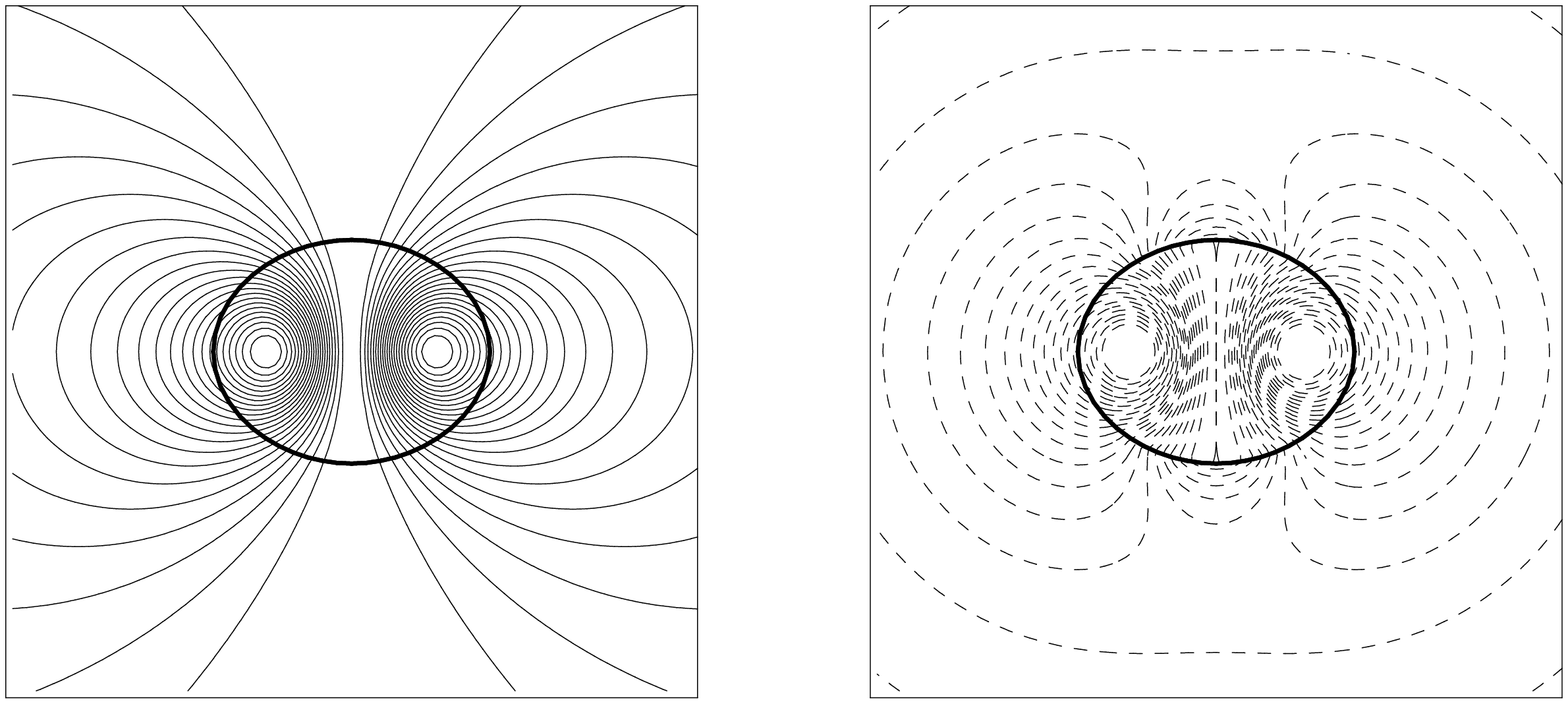,bbllx=100pt,bblly=18pt,bburx=440pt,bbury=774pt,angle=270,height=7cm,width=17.5cm}
\caption[]{\label{f:lignes,f=const}
Magnetic field lines (left) and electric isopotential lines $A_t={\rm const}$
(right) in the $(r,\theta)$ plane
for the configuration considered in Figs.~\ref{f:jphi,f=const},
\ref{f:B,f=const} and \ref{f:E,f=const}. The thick line denotes the
star's surface.}
\end{figure*}

\subsection{Structure of the electromagnetic field}

In order to consider different electric current distributions, concentrated
around the star's centre or not, we use different choices for the current
function
$f$. The most immediate choice is (\ref{e:f(x)=const}) ($f(x)=f_0={\rm
const}$).
The resulting electric current distribution is shown in
Fig.~\ref{f:jphi,f=const} for a Pol2 $M=3.37\ M_\odot$ model rotating at
$\Omega=3\times 10^3{\ \rm rad\, s}^{-1}$. $j^{(\phi)}$ has a maximum
at one third of the stellar radius.
The resulting magnetic field is shown in
Fig.~\ref{f:B,f=const} and \ref{f:lignes,f=const}. The magnetic field
amplitude has a maximum near the star's centre, the surface value being
approximatively four times lower.
The dominant dipolar structure of $\vec{B}$ clearly appears in
Fig.~\ref{f:lignes,f=const}, as if $\vec{B}$ was generated by a single current
loop located inside the star.
The induced electric field is represented in
Figs.~\ref{f:E,f=const} and \ref{f:lignes,f=const}.
Note on Fig.~\ref{f:E,f=const} the discontinuity in the component of $\vec{E}$
normal to the star's surface, due to surface charges.
The cusp across the surface in the component $E_{(\theta)}$ along the radius
$\theta=\pi/4$ is due to the discrepancy between $E_{(\theta)}$ and the
component of $\vec{E}$ tangential to the surface (resulting from the
non-spherical shape of the star), this latter
being perfectly smooth, as expected. The fact that $A_t$ is continuous but
not differentiable across the star's surface, because of the surface charges,
is clear on Fig.~\ref{f:lignes,f=const}.
Note also on Fig.~\ref{f:lignes,f=const} that, inside the star, the lines
$A_t={\rm const}$ coincide with the magnetic field lines, as expected from
the perfect conductor relation (\ref{e:conductparfait}) and the fact that
the magnetic field lines are the lines $A_\phi = {\rm const}$.
(cf. Sect.~\ref{s:def,not}).

\begin{figure}
\psfig{figure=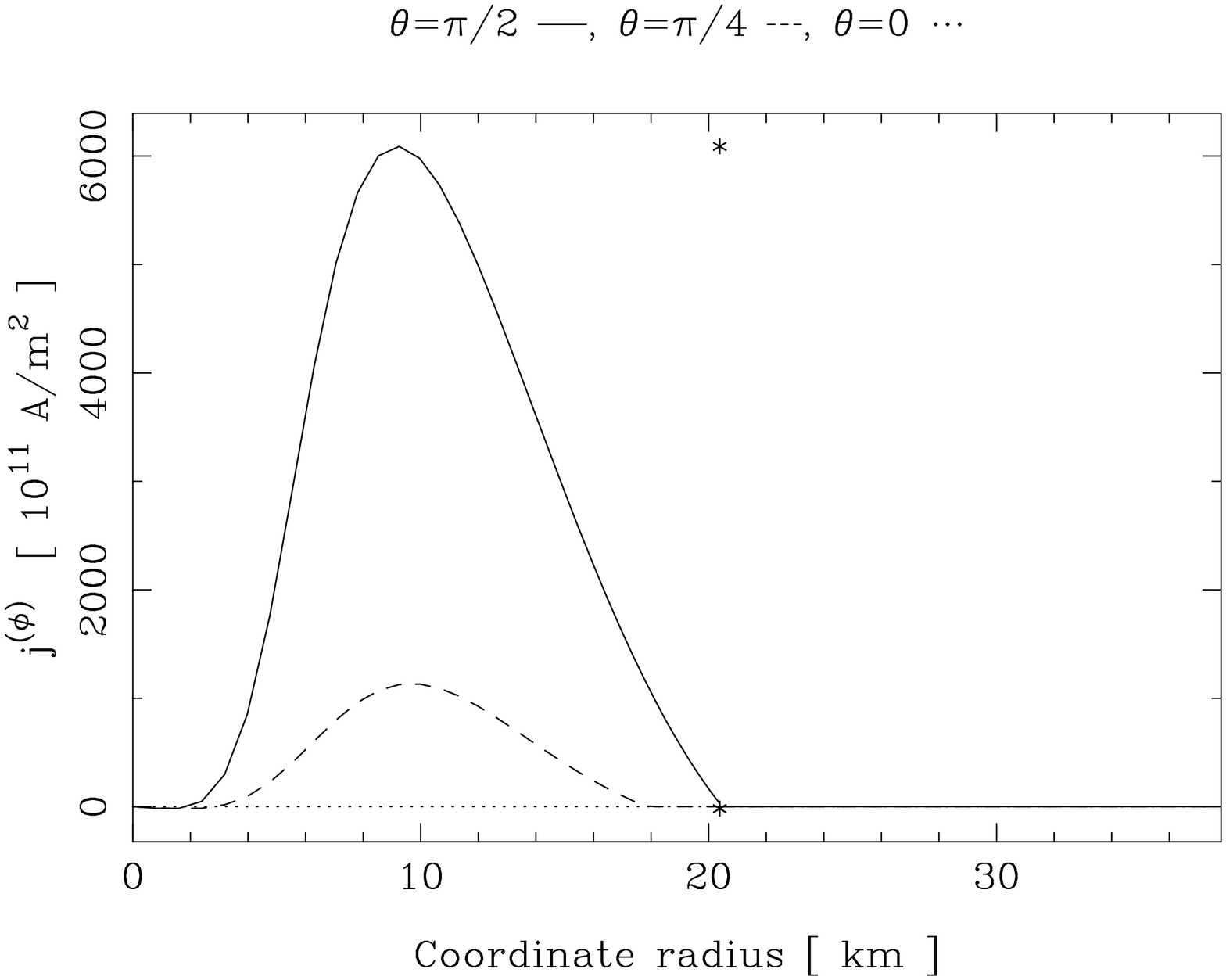,angle=270,height=7cm,width=8.5cm}
\caption[]{\label{f:jphi,f=1-Lorentz}
Electric current distribution induced by the choice $f(x) = 10^4 [ 1 - 1 /
( (3.3\times 10^{-3}\,  x)^2 + 1 ) ]  /R {\ \rm A\, m}^{-2}
 \rho_{\rm nuc}^{-1} {\rm c}^{-2}$ for a Pol2 EOS model, of central energy
density $e_{\rm c}=3.06 \ \rho_{\rm nuc}\, c^2$,
rotating at $\Omega= 3\times 10^3 {\ \rm rad\, s}^{-1}$. The notations are the
same as in Fig.~\ref{f:jphi,f=const}.}
\end{figure}

\begin{figure}
\psfig{figure=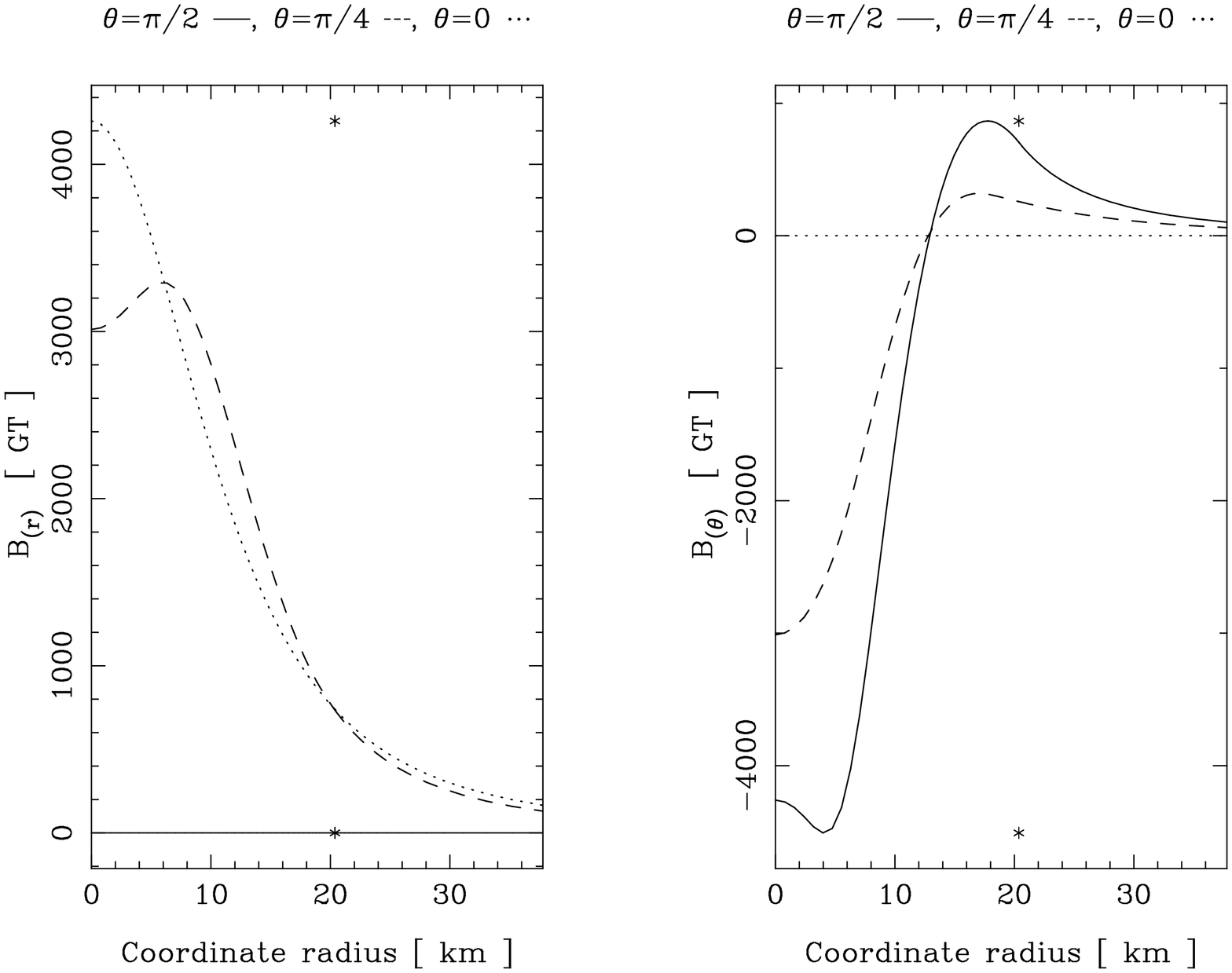,angle=270,height=7cm,width=8.5cm}
\caption[]{\label{f:B,f=1-Lorentz}
Components $B_{(r)}$ (left) and $B_{(\theta)}$ (right) of the magnetic field
generated by the electric current distribution of
Fig.~\ref{f:jphi,f=1-Lorentz}.
The notations are the same as in Fig.~\ref{f:jphi,f=const}.}
\end{figure}

\begin{figure}
\psfig{figure=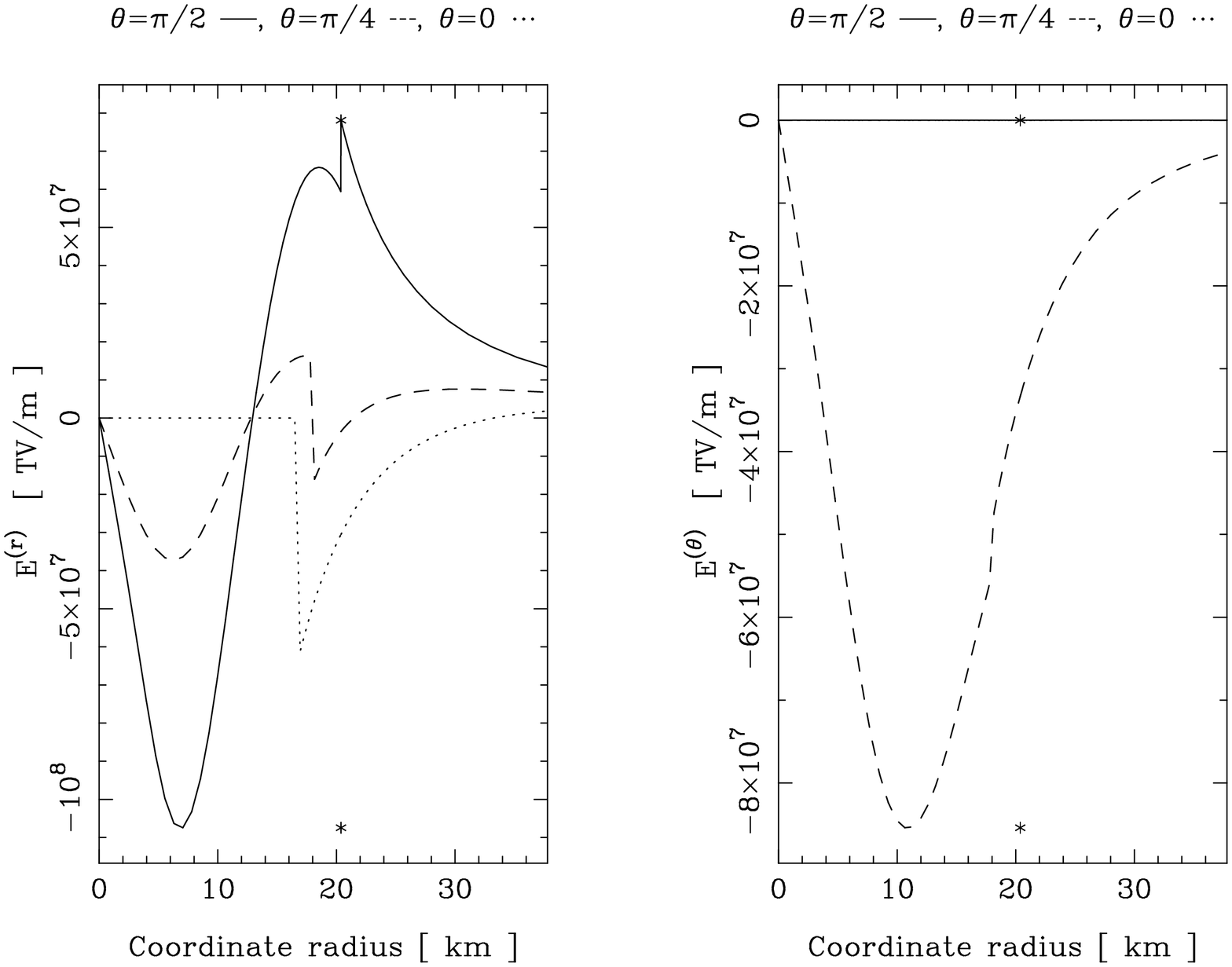,angle=270,height=7cm,width=8.5cm}
\caption[]{\label{f:E,f=1-Lorentz}
Components $E_{(r)}$ (left) and $E_{(\theta)}$ (right) of the electric field
corresponding to the configuration of Figs.~\ref{f:jphi,f=1-Lorentz} and
\ref{f:B,f=1-Lorentz}. The notations are the same as in
Fig.~\ref{f:jphi,f=const}.
On the right-hand figure, the dotted line is merged with the solid one.}
\end{figure}

In order to get current distributions different from the one described above,
we have considered the following forms of the current function, beside
the constant form (\ref{e:f(x)=const}),
\begin{eqnarray}
   \displaystyle f(x) & = & {\alpha \ov 1 + x}
	\ , \qquad x = {A_\phi\ov A_\phi^0} 	 \label{e:f=a/1+x} \\
   \displaystyle f(x) & = & \alpha \l( 1 - {1\ov (\beta x)^2 + 1} \r)
	\ , \qquad x = {A_\phi\ov A_\phi^0} \ ,	\label{e:f=1-Lorentz}
\end{eqnarray}
where $A_\phi^0 = 4\pi\times 10^4 \, R^3 \ {\rm T\, m}^{-1}$, $R$ being the
value of the coordinate $r$ at the equator.
One might think about other choices, such as $f(x)=\alpha x + \beta$,
$f(x) = \alpha x^2$, or $f(x)=\alpha/(1+x^2)$,
but with such functions, the iterative
procedure described in Sect.~\ref{s:resol} revealed not to converge,
resulting in a solution either cyclic or chaotic with respect to the number of
steps.
This behaviour is not due to some numerical instability but
results from the non-linear coupling between $j^\phi$ and $A_\phi$
via the Eqs.~(\ref{e:MaxAmp}) and (\ref{e:jphi=f(Aphi)}), which may
be written schematically as a unique equation $\Delta A_\phi = F(A_\phi)$.
Now it is well known that some sequences defined by an iterative
relation $u_{n+1} = F(u_{n})$ may exhibit a cyclic and/or chaotic
behaviour when $f$ is non-linear and not one-to-one
[take $F(x) = 4 \alpha (x - x^2)$, which generates cycles of increasing
order for $0.75 < \alpha < 0.892$ and chaotic sequences for $\alpha > 0.892$].
For this reason, we did not consider the current functions mentionned above
in our study.

The choice (\ref{e:f=a/1+x}) leads to a current distribution slightly different
from that corresponding of $f(x)={\rm const.}$ (Fig.~\ref{f:jphi,f=const}),
being simply more concentrated towards the star's centre, especially in the
direction of the equatorial plane. The resulting electromagnetic field
is very similar to that of Figs.~\ref{f:B,f=const}-\ref{f:lignes,f=const}.

The choice (\ref{e:f=1-Lorentz}) leads to a current distribution
which is represented in Fig.~\ref{f:jphi,f=1-Lorentz}. There is no current
around the centre and the maximum of the distribution is reached at
almost half of the stellar radius.
The resulting magnetic field is shown in Fig.~\ref{f:B,f=1-Lorentz}.
The difference with that generated by $f(x)={\rm const}$
(cf. Fig.~\ref{f:B,f=const}) is not important; simply the maximum
of $\vec{B}$ is achieved apart from the star's centre (except near the
rotation axis). The corresponding electric field is shown in
Fig.~\ref{f:E,f=1-Lorentz}. It has more or less the same structure as
that of Fig.~\ref{f:E,f=const}.

\begin{figure}
\psfig{figure=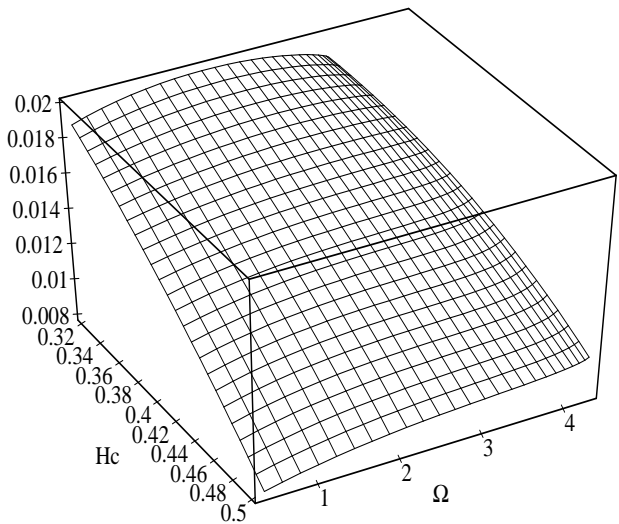,bbllx=1.5in,bblly=1in,bburx=7in,bbury=9in,angle=270,height=7cm,width=8.5cm}
\caption[]{\label{f:diff:Mbar}
Difference between the baryonic mass [in $M_\odot$] of configurations
with a magnetic dipole moment ${\cal M} = 1.5\ 10^{32} {\ \rm A\, m}^2$
and the baryonic mass of configurations without magnetic field,
as a function of the central log-enthalpy $H_{\rm c}$ and the angular
velocity $\Omega$ [in $10^3\ {\rm rad\, s}^{-1}$], for the Pol2 EOS.}
\end{figure}

\begin{figure}
\psfig{figure=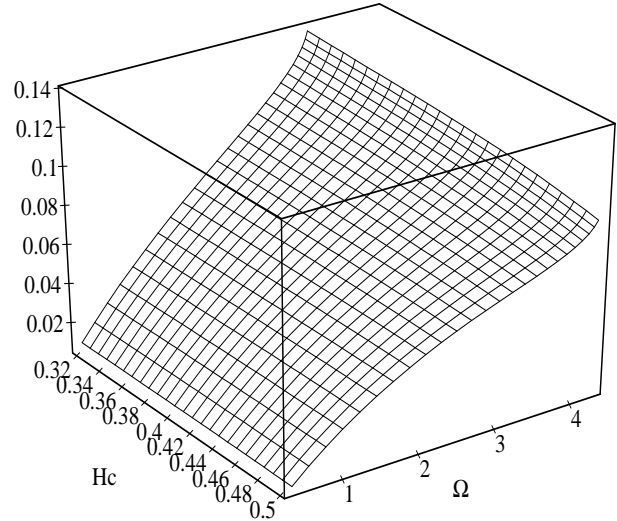,bbllx=1.5in,bblly=1in,bburx=7in,bbury=9in,angle=270,height=7cm,width=8.5cm}
\caption[]{\label{f:diff:J}
Same as Fig.~\ref{f:diff:Mbar} but for the total angular momentum $J$ [in
$M_\odot \, c \, {\rm km}$].}
\end{figure}

\begin{figure}
\psfig{figure=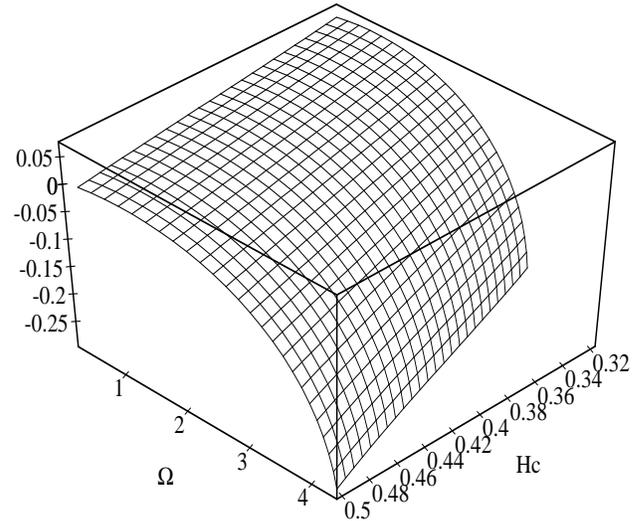,bbllx=1.5in,bblly=1in,bburx=7in,bbury=9in,angle=270,height=7cm,width=8.5cm}
\caption[]{\label{f:diff:Rcirc}
Same as Fig.~\ref{f:diff:Mbar} but for the equatorial circumferential radius
$R_{\rm circ}$ [in km].}
\end{figure}

\begin{figure}
\psfig{figure=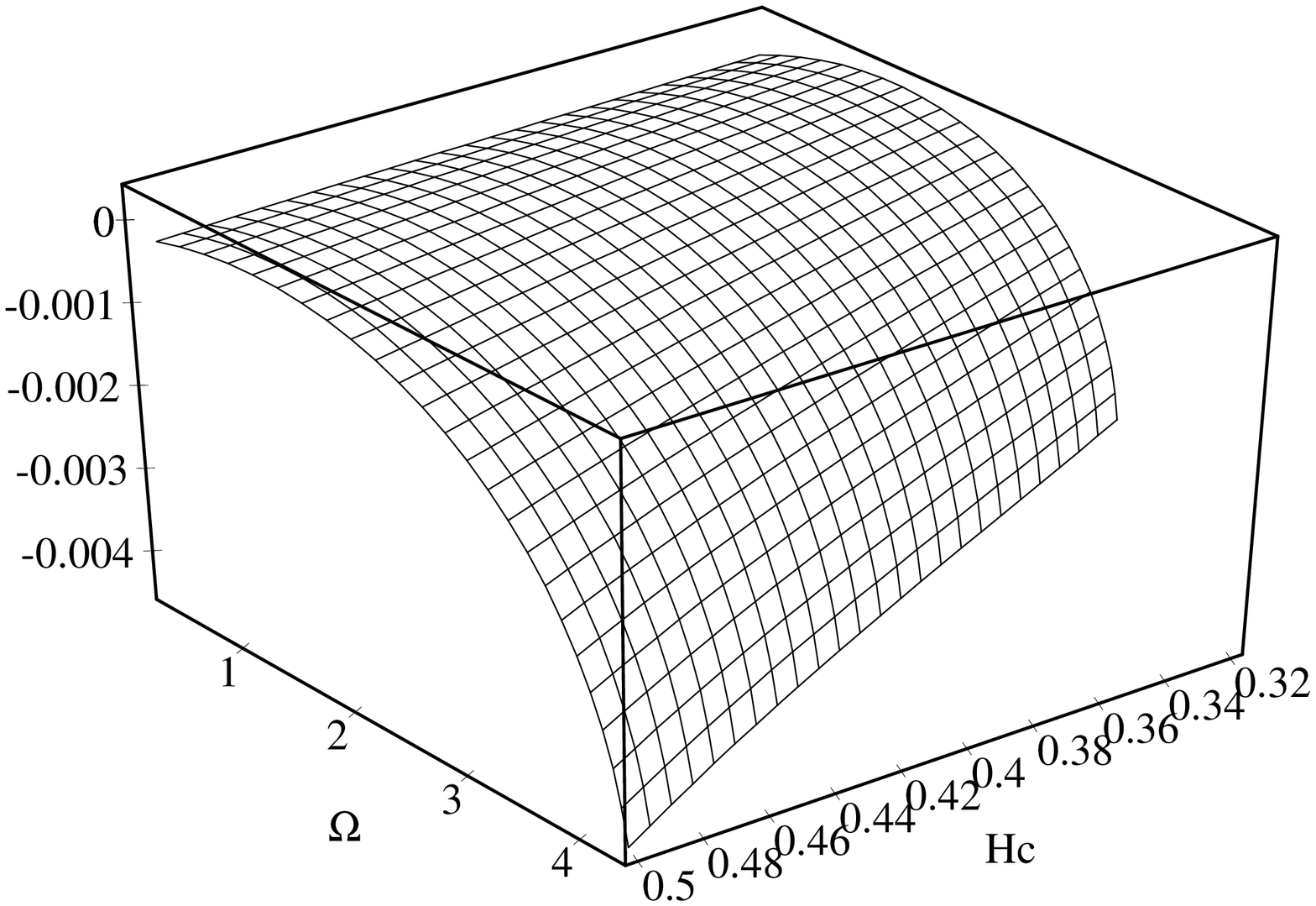,bbllx=1.5in,bblly=1in,bburx=7in,bbury=9in,angle=270,height=7cm,width=8.5cm}
\caption[]{\label{f:diff:U}
Same as Fig.~\ref{f:diff:Mbar} but for the equatorial fluid velocity as
measured by the locally non-rotating observer, $U_{\rm eq}$ [in units of $c$].}
\end{figure}

\subsection{Differences between magnetized and non-magnetized configurations}

For a given EOS and a given choice of the current function,
we have explored the plane $(H_{\rm c},\Omega)$ by computing
around $14\times 14 = 196$ models, with and without electromagnetic
field. We present in this section the results concerning the effect of the
electromagnetic field in the $(H_{\rm c},\Omega)$ space for the
particular case of the Pol2 EOS and the form (\ref{e:f(x)=const}) of the
current function, the constant $f_0$ being adjusted in each model  in
order to lead to the same magnetic dipole moment
${\cal M} = 1.5\ 10^{32} {\ \rm A\, m}^2$, so that the magnetized
configurations may be considered as equivalent from a magnetic point of
view for an observer sufficiently far from the star.

A first effect of the magnetic field is shown on
Fig.~\ref{f:diff:Mbar}: it increases the
baryonic number at fixed $H_{\rm c}$ and $\Omega$. This means that Lorentz
forces, thanks to their centrifugal effect (cf. Sect.~\ref{s:static:deform}),
help the star to support more baryons at the same central density and same
rotation rate. It appears on Fig.~\ref{f:diff:Mbar} that this effect
is rather insensitive to $\Omega$, reflecting that this is the
magnetic field which contributes the most to the Lorentz force, not the
electric field.
It can be verified that, as in the static case, the
centrifugal effect of Lorentz forces enhances the oblateness of the star.

A second effect of the magnetic field is to increase the total angular
momentum $J$, as shown in Fig.~\ref{f:diff:J}, more or less linearly with
respect to $\Omega$. This augmentation is rather significative and can reach
more
than $1.5\%$ of the zero magnetic field value. The redistribution of baryons
inside the star induced by Lorentz forces
is responsible for most of the increase, well before the contribution to $J$
of the electromagnetic field intrinsic angular momentum.

The variation of the star's equatorial radius $R_{\rm circ}$ (defined as
the star's equatorial circumference as measured by the observer ${\cal O}_0$
divided by $2\pi$) is depicted in Fig.~\ref{f:diff:Rcirc}. For small
angular velocities, the radius increases, in the continuation of the
static case where it has been seen that
Lorentz forces stretch the star out (Sect.~\ref{s:static:deform}).
But for larger angular velocities the radius decreases instead. This
may be explained in the following way. At large rotation rates the stellar
equilibrium is governed by the inertial centrifugal forces.
Now these forces increase with the radius and are rather important at the
periphery of the star. On the contrary, the Lorentz forces are important
near the centre and support there the massive nucleus of the star, so
that for the same central density and rotation rate, the star can have a
smaller radius to maintain the equilibrium, which is energetically favorable
in so far as the peripheric inertial forces are near the shedding limit,
contrary
to the low rotation velocity case.
Due to this reduction of the star's radius, the equatorial fluid velocity
as measured by the observer ${\cal O}_0$ decrease, as it can be verified on
Fig.~\ref{f:diff:U}.

\begin{table*}
\caption[]{\label{t:rot,Mmax}
Neutron star configurations in the rotating case. For each EOS, the
first two lines are maximum mass configurations at fixed $\cal M$.
$\Omega$ is the angular velocity; for the first two lines of each EOS $\Omega$
is
the Keplerian velocity $\Omega_{\rm K}$.
$P={2\pi /\Omega}$ is the rotation period,
$R_{\rm circ}$ the equatorial circumferential radius,
$J$ the angular momentum, $Q$ the quadrupole moment as defined in
Salgado et al. 1994a,
$U_{\rm eq}$ the fluid velocity at the equator as measured by
a locally non-rotating observer and $N^\phi_{\rm c}$ the ``dragging of
the inertial frames'' at the star's centre.
The remaining symbols are defined in the caption of Table~\ref{t:static,Mmax}.
We use
$G=6.6726\ 10^{-11}\ {\rm m}^3{\rm kg}^{-1}{\rm s}^{-2}$,
$c=2.997925\ 10^8{\ \rm m\, s}^{-1}$ and
$1\ M_\odot=1.989\ 10^{30}\ {\rm kg}$.}
\begin{flushleft}
\begin{tabular}{rrrrrrrrrrrr}
\hline\noalign{\smallskip}
 $\displaystyle {{\rm EOS} \atop \ }$  &
 $\displaystyle {{\cal M} \atop [10^{32}\, {\rm A\, m}^2]}$  &
 $\displaystyle {B_{\rm c} \atop [10^{3}\, {\rm  GT}]}$ &
 $\displaystyle {B_{\rm pole} \atop [10^{3}\, {\rm  GT}]}$ &
 $\displaystyle {H_{\rm c} \atop \ }$ &
 $\displaystyle {e_{\rm c}\atop [\rho_{\rm nuc} c^2]}$  &
 $\displaystyle {p_{\rm mg,c} \ov p_{\rm fl,c}}$  &
 $\displaystyle {\Omega\atop [10^4\, {\rm s}^{-1}]}$ &
 $\displaystyle {P\atop [{\rm ms}]}$  &
 $\displaystyle {M\atop [M_\odot]}$ &
 $\displaystyle {{\cal B}\atop [M_\odot]}$  &
 $\displaystyle {E_{\rm bind}\atop [m_{\rm B} c^2]}$ \\
 \noalign{\smallskip}
\hline\noalign{\smallskip}
Pol2     &0&0&0& 0.432 &  3.43& 0 & 0.398 & 1.58 & 3.63 &3.99&-0.0897\\
         &1.95&22.6&5.42& 0.432 &  3.43&0.02 & 0.404 & 1.56 & 3.68
&4.04&-0.0880\\
        &3.98&41.9&11.3& 0.400 &  3.06&0.08 & 0.389 & 1.62 & 3.76
&4.11&-0.0852\\
 \noalign{\smallskip}
\hline\noalign{\smallskip}
PandN    &0&0&0& 0.668 & 21.7&0 & 1.29 & 0.488 & 1.93 &2.23&-0.1313\\
         &0.24&54.6&15.2& 0.664 & 21.5&0.01 & 1.29 & 0.487 & 1.94
&2.23&-0.1297\\
        &1.07&251.9&88.7& 0.600 & 19.1&0.23 & 1.16 & 0.543 & 1.97
&2.20&-0.1071\\
 \noalign{\smallskip}
\hline\noalign{\smallskip}
BJI      &0&0&0& 0.628&16.3&0&1.07&0.588&2.15 &2.46&-0.1240\\
         &0.50&68.0&19.0 &0.619&16.0&0.02&1.08&0.584&2.18 &2.48&-0.1207\\
        &1.45&172.5&60.2 &0.500&12.5&0.21&0.896&0.701&2.19&2.45&-0.1066\\
\noalign{\smallskip}
\hline
\end{tabular}
\begin{tabular}{rrrrrrrr}
\hline\noalign{\smallskip}
 $\displaystyle {{\rm EOS} \atop \ }$  &
 $\displaystyle {R_{\rm circ}\atop [{\rm km}]}$  &
 $\displaystyle {c J \over G\, M^2}$ &
 $\displaystyle {Q\over M\, R_{\rm circ}^2}$ &
 $\displaystyle {U_{\rm eq} \over c}$ &
 $\displaystyle {N^\phi_{\rm c} \over \Omega}$ &
 $\displaystyle {{\rm GRV2}\atop \ }$  &
 $\displaystyle {{\rm GRV3}\atop \ }$  \\
\noalign{\smallskip}
\hline\noalign{\smallskip}
 Pol2 &31.0 & 0.570  & 0.0239& 0.468 & 0.648 & 6E-06 & 1E-05 \\
      &31.1 & 0.580  & 0.0256& 0.475 & 0.657 & 9E-03 & 1E-03 \\
      &29.3 & 0.589  & 0.0344& 0.427 & 0.658 & 4E-03 & 4E-04 \\
 \noalign{\smallskip}
\hline\noalign{\smallskip}
 PandN &11.4 & 0.641 & 0.0320& 0.576 & 0.805 & 1E-04 & 7E-05 \\
       &11.4 & 0.646 & 0.0325& 0.579 & 0.807 & 7E-03 & 6E-04 \\
       &9.9  & 0.593 & 0.0444& 0.431 & 0.833 & 6E-03 & 7E-04 \\
 \noalign{\smallskip}
\hline\noalign{\smallskip}
 BJI   &13.4 & 0.626 & 0.0298& 0.559 & 0.784 & 4E-05 & 9E-05 \\
       &13.4 & 0.638 & 0.0318& 0.566 & 0.789 & 5E-03 & 4E-04 \\
       &12.1 & 0.575 & 0.0433& 0.407 & 0.780 & 5E-03 & 3E-04 \\
\noalign{\smallskip}
\hline
\end{tabular}
\end{flushleft}
\end{table*}

\begin{figure*}
\psfig{figure=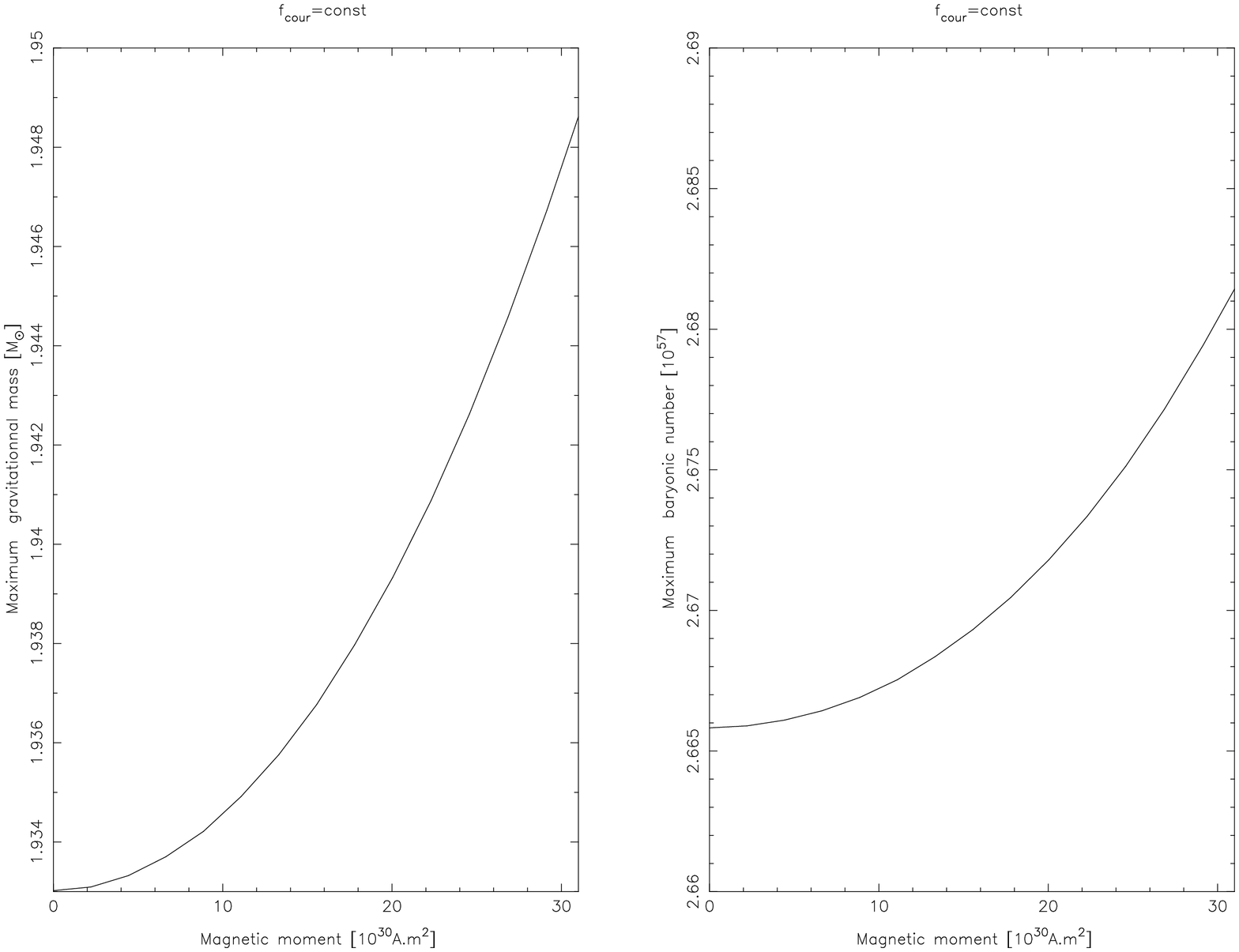,angle=270,height=8cm,width=17.5cm}
\caption[]{\label{f:rot,Mmax,PandN}
Maximum gravitational mass (left) and maximum baryon number
(right) along rotating sequences at fixed magnetic dipole moment
as a function of the magnetic dipole moment for the PandN EOS.}
\end{figure*}

\begin{figure}
\psfig{figure=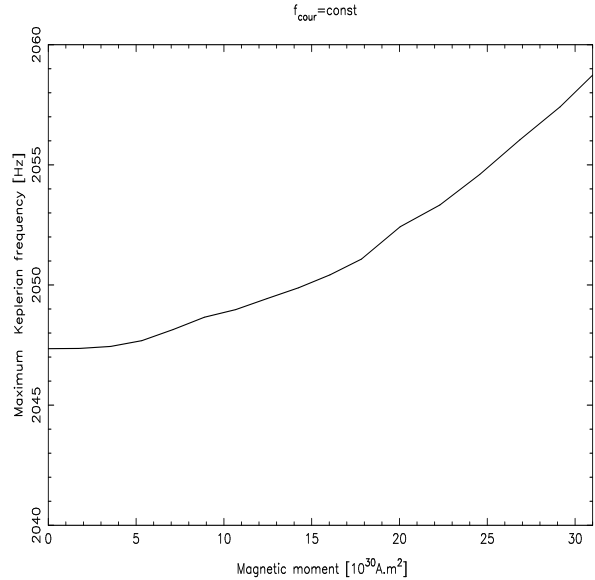,angle=270,height=8cm,width=8.5cm}
\caption[]{\label{f:rot,Omega_max,PandN}
Maximum Keplerian frequency $\Omega_{\rm K}/2\pi$ along rotating sequences at
fixed magnetic dipole moment
as a function of the magnetic dipole moment for the PandN EOS.}
\end{figure}

\begin{figure*}
\psfig{figure=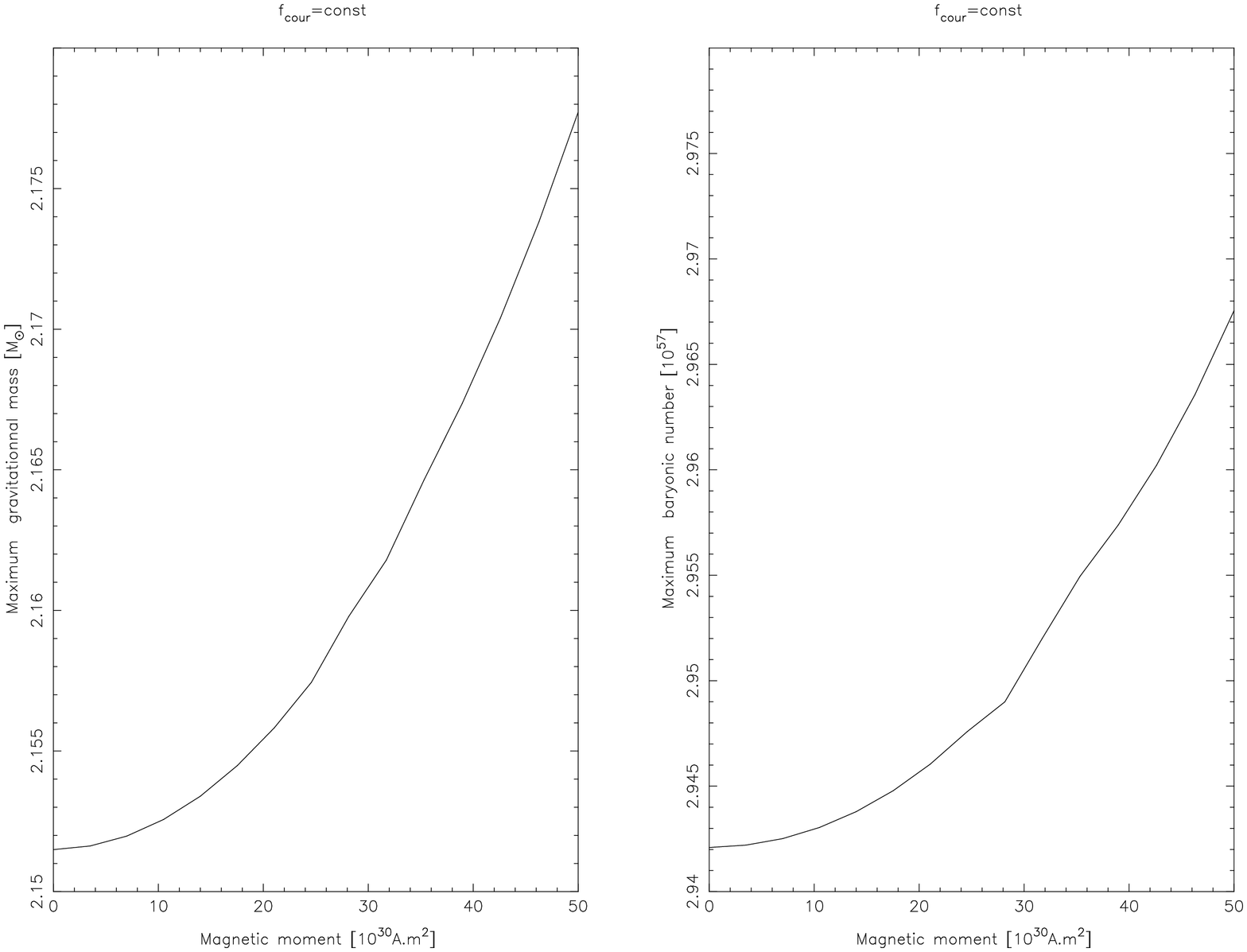,angle=270,height=8cm,width=17.5cm}
\caption[]{\label{f:rot,Mmax,BJI}
Same as Fig.~\ref{f:rot,Mmax,PandN} but for the BJI EOS.}
\end{figure*}

\begin{figure}
\psfig{figure=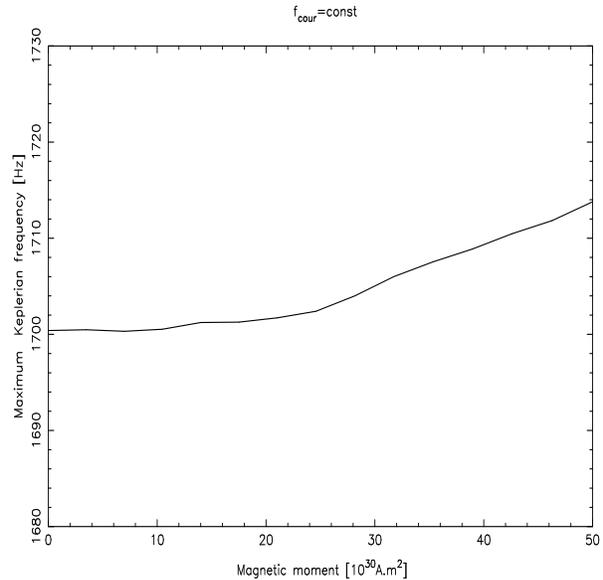,angle=270,height=8cm,width=8.5cm}
\caption[]{\label{f:rot,Omega_max,BJI}
Same as Fig.~\ref{f:rot,Omega_max,PandN} but for the BJI EOS.}
\end{figure}

\subsection{Effect on $M_{\rm max}$ and $\Omega_{\rm max}$}

Table~\ref{t:rot,Mmax} gives (i) maximum mass rotating configurations
without any magnetic field (first line), (ii) along a sequence of constant
magnetic dipole moment $\cal M$ (second line)
and (iii) rotating configurations with magnetic fields close to the
maximum value (third line), for the EOS Pol2, PandN and
BJI.

As in the static case, the maximum gravitational mass along
sequences at fixed $\cal M$, $M_{\rm max}$, increases with $\cal M$
(cf. Figs.~\ref{f:rot,Mmax,PandN} and \ref{f:rot,Mmax,BJI}).
A difference with the static case is that the maximum baryon number is
an increasing function of $\cal M$ for all EOS, including BJI (compare
Figs.~\ref{f:static,Mmax,BJI} and \ref{f:rot,Mmax,BJI}).

For a given central-enthalpy and magnetic dipole moment, the Keplerian angular
velocity $\Omega_{\rm K}$ is the value of $\Omega$ above which no stationary
solution exists, the star being at the break up limit under the effect
of centrifugal inertial forces. The maximum Keplerian velocity along a
sequence at fixed $\cal M$ is reached at, or very close to, the maximum mass
configuration. As seen from Figs.~\ref{f:rot,Omega_max,PandN} and
\ref{f:rot,Omega_max,BJI}, the maximum $\Omega_{\rm K}$ increases with
$\cal M$.

\begin{figure*}
\psfig{figure=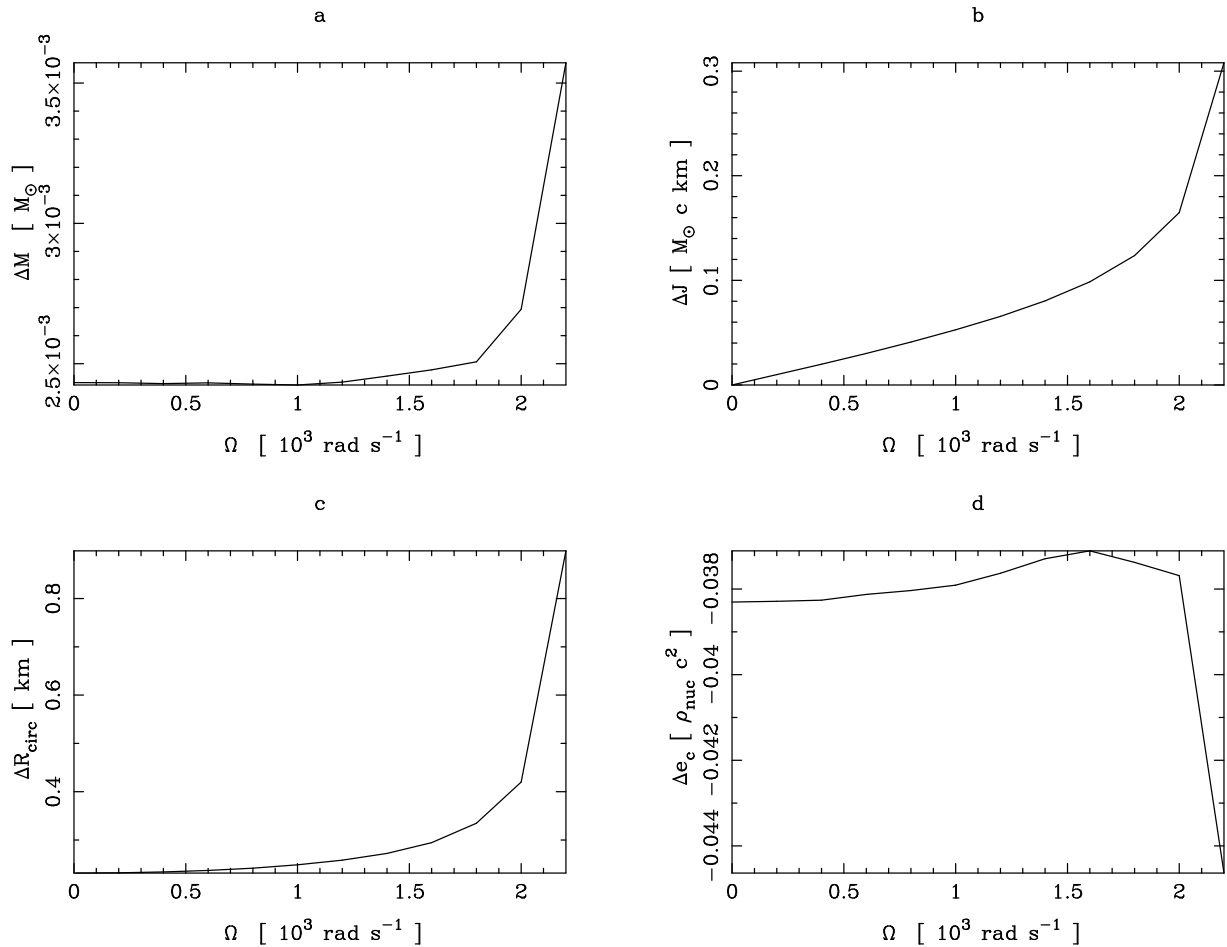,angle=270,height=13cm,width=17.5cm}
\caption[]{\label{f:seqB=3.0,Pol2}
Differences between configurations without magnetic field and with a
magnetic dipole moment ${\cal M} = 1.5\ 10^{32} {\ \rm A\, m}^2$ along
a sequence at constant baryonic mass ${\cal B} = 3.00 \ M_\odot$ for the Pol2
EOS and the current function $f(x)={\rm const}$: (a) gravitational
mass difference $\Delta M$; (b) angular momentum difference $\Delta J$;
(c) circumferential radius difference $\Delta R_{\rm circ}$; (d) central energy
density difference $\Delta e_{\rm c}$. The path along the sequence is
parametrized by the angular velocity $\Omega$.}
\end{figure*}

\section{Magnetized sequences at constant baryon number} \label{s:const,bar}

Constant baryon number sequences may represent the time evolution of a
neutron star and have been investigated in recent studies of non-magnetized
rotating neutron stars (Cook et al. 1994b, Salgado et al. 1994a,b).

We analyze in this section the behaviour of constant baryon number sequences
at a fixed magnetic dipole moment ${\cal M} = 1.5\ 10^{32} {\ \rm A\, m}^2$,
by stressing the differences with the non-magnetized case.

\subsection{Normal sequence}

A {\em normal} sequence is a sequence with a baryon number lower
than the maximum one supported by non-rotating configurations. We consider a
sequence
at baryonic mass ${\cal B} = 3.00 \ M_\odot$, built with the
Pol2 EOS and the choice (\ref{e:f(x)=const}) for the current function $f$.
Let us first mention that the choice ${\cal M} = 1.5\ 10^{32} {\ \rm A\, m}^2$
corresponds to highly magnetized objects, the polar value of the magnetic
field varying from $B=1.4\ 10^3 {\ \rm GT}$ ($\Omega=0$) to
$B=5.0\ 10^2 {\ \rm GT}$ ($\Omega=\Omega_{\rm K}$) along the sequence.
The difference between various characteristic quantities of magnetized
and non-magnetized stars of the sequence are represented in
Fig.~\ref{f:seqB=3.0,Pol2}. According to Fig.~\ref{f:seqB=3.0,Pol2}a
the magnetic field increases the gravitational mass, but only by a small
amount (for the chosen value of $\cal M$), around one thousandth.
The angular momentum increase is more
substantial, being of the order of $3\,\%$ (Fig.~\ref{f:seqB=3.0,Pol2}b).
The star's radius increases under the centrifugal effect of the Lorentz forces
by roughly $5\, \%$ (Fig.~\ref{f:seqB=3.0,Pol2}c) and consequently the central
energy density decreases, by roughly $4\, \%$ (Fig.~\ref{f:seqB=3.0,Pol2}c).
For this normal sequence, the response of the global quantities to the
magnetic field is the same as in the static case (cf Sect.~\ref{s:static}).

We have also constructed normal sequences with the PandN EOS. Their behaviour
is very similar to that described above.

\begin{figure*}
\psfig{figure=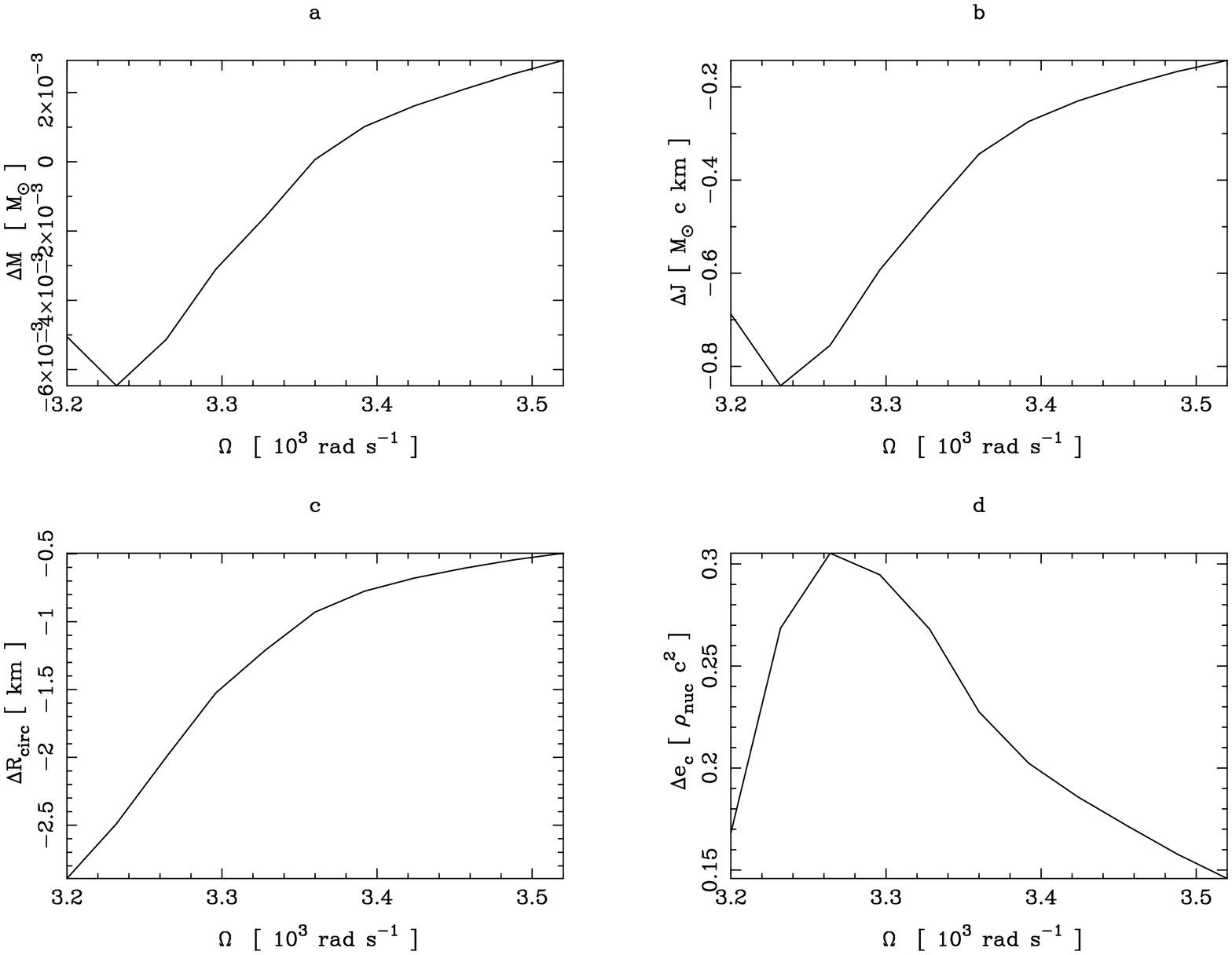,angle=270,height=13cm,width=17.5cm}
\caption[]{\label{f:seqB=3.8,Pol2}
Same as Fig.~\ref{f:seqB=3.0,Pol2} but for a supramassive sequence
at ${\cal B} = 3.80 \, M_\odot$.}
\end{figure*}

\subsection{Supramassive sequence}

A {\em supramassive} sequence is a sequence with a
baryon number in excess of the maximum value that can be supported in the
absence of rotation. We consider a supramassive sequence which in addition
presents the phenomenon of {\em spin-up by angular
momentum loss} (Cook et al. 1994a): the ${\cal B}=3.80\  M_\odot$ sequence
built on the Pol2 EOS. The fact that along this sequence a decrease of
angular momentum leads to an increase of angular
velocity is demonstrated in Fig.~7 of Salgado et al. (1994a).
The sequence begins at the angular velocity
$\Omega=3.20\, 10^3 {\ \rm rad\, s}^{-1}$ and ends at the stability
limit $\Omega=3.52\, 10^3 {\ \rm rad\, s}^{-1}$. The electromagnetic
effects are reported in Fig.~\ref{f:seqB=3.8,Pol2} and appear to differ
from those of the normal sequence. The gravitational mass decreases for
the lowest angular velocities and increases for the highest ones
(Fig.~\ref{f:seqB=3.8,Pol2}a), whereas the angular momentum decreases
all the time (Fig.~\ref{f:seqB=3.8,Pol2}b). This decrease in angular
momentum can be attributed mainly to the star's radius decrease
(Fig.~\ref{f:seqB=3.8,Pol2}c). This latter behaviour contrasts with
the normal sequence case (cf. Fig.~\ref{f:seqB=3.0,Pol2}c) and can be
explained as follows. Near the minimum angular velocity
($\Omega=3.20\, 10^3 {\ \rm rad\, s}^{-1}$), the centrifugal forces
take their origin, not in a high rotation rate, but in a large spatial
extension. The Lorentz forces play then a greater role in the equilibrium
of the star since they are mainly localised in the star's centre
and are more efficient in supporting the star than the centrifugal forces,
which are localised at the star's periphery.
The replacement of the centrifugal
forces by the Lorentz ones in balancing gravity leads
to a smaller stellar radius.
This can be illustrated by the following example: a star
(built upon the Pol2 EOS) of baryon mass $3.77\, M_\odot$ and central density
$3.1\, \rho_{\rm nuc}$
has a circumferential radius of $26.5$ km if it is supported by rotation
($\Omega=3.3\times 10^3 \, {\rm rad\, s}^{-1}$) and only
$21.2$ km if it is static and supported by the magnetic field
(${\cal M} = 7.02\times 10^{32} \, {\rm A\, m}^2$), for the same number
of baryons.
This phenomenon explains the negative value of
$\Delta R_{\rm circ}$ in Fig.~\ref{f:seqB=3.8,Pol2}c
which reaches $-9\, \%$ at $\Omega=3.20\, 10^3 {\ \rm rad\, s}^{-1}$.
When $\Omega$ increases the centrifugal forces become more efficient
and the decrease of the star's radius with respect to the
non-magnetized case becomes less pronounced.
Due to the contraction of the star, the central density increases
(Fig.~\ref{f:seqB=3.8,Pol2}d). According to this explanation, the same
phenomenon should exist in the Newtonian case, but we have not conducted
a systematic study of Newtonian configurations.

\section{Conclusion} \label{s:concl}

We have extended an existing numerical code for computing perfect fluid
rotating neutron stars in general relativity (BGSM, Salgado et al. 1994a,b)
to include the electromagnetic field. This latter is calculated by solving
the relativistic Maxwell equations with an electric current distribution which
is compatible with the star's equilibrium (i.e. the Lorentz force acting
on the conducting fluid shall be the gradient of some scalar in order to
balance
gravity and the inertial centrifugal force).
In order to preserve the stationarity, axisymmetry and circularity properties
of spacetime, we consider only axisymmetric poloidal magnetic fields.
The equations are numerically
solved by means of a pseudo-spectral technique which results in a high
accuracy as tests on simple electromagnetic configurations (for which an
analytical solution is available) have shown: the relative error on the
electromagnetic field is of the order of $10^{-5}$ inside the star and
$10^{-9}$ outside it. The part of the code relative to the
deformation of the star by Lorentz forces has been tested by comparison
with Ferraro's analytical solution in the Newtonian case.
The fact that the numerical output is a solution of Einstein equations
has been tested by two virial identities: GRV2 and GRV3.

We have then used the code to investigate the effect of the magnetic
field on rotating neutron stars. For this purpose we considered magnetic
field amplitude ranging from zero up to huge values, of the order of
$10^5 {\ \rm GT}$, which is ten thousand times bigger than the highest measured
values at the pulsars surfaces and is the value for which the magnetic
pressure equals (with an opposite sign along the symmetry axis) the fluid
pressure near the centre of the star. Let us note that such enormous
magnetic fields are expected to decay on very short time scales via the
mecanism of ambipolar diffusion, as investigated by Haensel et al. (1991),
Goldreich \& Reisenegger (1992) and Urpin \& Shalybkov (1995),  which is very
efficient when the electric current is perpendicular to $\vec{B}$, as in the
present case. The decay time scale for non-superconducting matter computed by
the above authors is
$10^2$ yr for $B\sim 10^4 {\ \rm GT}$ and $10^6$ yr for
$B\sim 1 {\ \rm GT}$.

According to our study, the
influence of the magnetic field on the star's structure is mostly due to
Lorentz forces and not to the gravitational field generated by the
electromagnetic stress-energy. This may be understood once it has been
realized that a magnetic field of $10^5{\ \rm GT}$ has an energy density
of $0.25\ \rho_{\rm nuc} c^2$, whereas the matter density at the
centre of neutron stars is between $1$ and $10\ \rho_{\rm nuc}$.
Although the electromagnetic energy is much lower than
the fluid mass-energy, the deformation of the star can be as dramatic
as that of Fig.~\ref{f:static,isoener,Mmax} because of the {\em anisotropic}
character of the magnetic pressure, just as the anisotropic centrifugal
forces can highly deform the star in the rotating case though the kinetic
energy is much lower than the fluid rest mass.

In static and slowly rotating cases, Lorentz forces stretch the star away
from the symmetry axis. The deformation is appreciable only for
$B>10^2{\ \rm GT}$. In highly relativistic situations (supramassive sequences),
the effect of the magnetic field is instead to reduce the star's
equatorial radius (at fixed baryon number).

The maximum poloidal magnetic field supported by neutron stars
has a polar value between $4\times 10^4$ and $1.5\times 10^5{\ \rm  GT}$
depending upon the
EOS and the rotation state of the star. Let us recall that the magnetic field
at the star's centre is two to four times higher than at the poles.

The impact of the magnetic field on the maximum mass of neutron stars is
very limited for magnetic fields of the order of 1 GT, whereas it is
important for the magnetic fields near the maximum value
($\sim 10^5{\ \rm  GT}$):
in the static case, $M_{\rm max}$ is increased by $13\%$ to $29\%$ ---
depending upon the EOS --- with respect to non-magnetized configurations.
In fact, the magnetic field reveals to be more efficient in increasing
$M_{\rm max}$ than the rotation, except for the EOS HKP, where the
maximum mass in rotation (without any magnetic field) is $21.2 \%$
higher than $M_{\rm max}$ for stationary configurations
(Salgado et al. 1994a), whereas
the static magnetized $M_{\rm max}$ is only $13.3\%$ higher. For the
PandN EOS, the $M_{\rm max}$ increase by both mecanisms are similar
($\sim 16 \%$), whereas for the Pol2, BJI and Diaz II EOS, the static
magnetized
$M_{\rm max}$ lies above the rotating non-magnetized $M_{\rm max}$.

In the future, we plan to study the stability of the magnetized configurations
presented in this article. Two types of instabilities may be expected to
occur for high values of $\vec{B}$: (i) a pure electromagnetic
instability towards another electric current - magnetic field distribution
(of lower energy) and (ii) a non-axisymmetric instability resulting in a
triaxial stellar equilibirum shape, which would be the magnetic analog of
the transition from the MacLaurin spheroids to the tri-axial Jacobi ellipsoids
for high rotational velocities of Newtonian incompressible bodies.

\begin{acknowledgements}
We thank Pawel Haensel for his careful reading of the manuscript and
many remarks about the physics of the magnetic field in neutron star
interiors. We are grateful to
the anonymous referee for useful comments that helped us to improve the paper.
The numerical computations have been performed on Silicon Graphics workstations
purchased thanks to the support of the SPM department of the CNRS and the
Institut National des Sciences de l'Univers.
\end{acknowledgements}

\end{document}